\documentclass[sigconf]{acmart}

\usepackage[english]{babel}
\usepackage{booktabs}
\usepackage{xspace}
\usepackage{hyperref}

\usepackage[capitalise,nameinlink,noabbrev]{cleveref}
\crefname{appendix}{Appx.}{Appx.}
\crefname{section}{\S}{\S{}}
\crefname{figure}{Fig.}{Fig.}
\crefname{table}{Tab.}{Tab.}
\usepackage{multirow}
\usepackage{makecell}
\usepackage[dvipsnames]{xcolor}

\usepackage{float} 
\usepackage{subfigure}

\newcommand{\projname}{\textsc{ACOS}\xspace}
\newcommand{\longprojname}{Arrays of Cheap Optical Switches\xspace}
\title{\projname: \longprojname}

\author{Daniel Amir}
\affiliation{%
  \institution{Technion}
}
\author{Ori Cohen}
\affiliation{%
  \institution{Technion}
}
\author{Jakob Krebs}
\affiliation{%
  \institution{Technion}
}
\author{Mark Silberstein}
\affiliation{%
  \institution{Technion}
}

\newcommand{\by}{$\times$}
\newcommand{\mypara}[1]{\noindent {\bf#1.}}
\newcommand{\alltoallv}[0]{AlltoAll(V)\xspace}
\newcommand{\alltoall}[0]{AlltoAll\xspace}
\newcommand{\allreduce}[0]{AllReduce\xspace}
\newcommand{\allgather}[0]{AllGather\xspace}
\newcommand{\reducescatter}[0]{ReduceScatter\xspace}

\renewcommand{\paragraph}[1]{\mypara{#1}}

\renewcommand\footnotetextcopyrightpermission[1]{} 
\setcopyright{none}

\acmDOI{}

\acmISBN{}

\acmConference[Arxiv]{}
\acmYear{2026}

\acmPrice{}

\usepackage{ifthen}
\newbool{public_version}
\setbool{public_version}{true}

\ifbool{public_version}{
}{

}

\begin{document}

\sloppy

\begin{abstract}
    Machine learning training places immense demands on cluster networks, motivating specialized architectures and co-design with parallelization strategies.
Recent designs incorporating optical circuit switches (OCSes) are promising, offering improved cost, power efficiency, and long-term bandwidth scaling than packet switches. However, most existing approaches rely on costly high-radix OCSes and/or combine them with packet switches to achieve competitive performance at scale.
Unfortunately, high-radix OCSes are both expensive and slow to reconfigure, limiting both scalability and performance.

We propose \emph{\longprojname} (\projname), which bring application co-design directly to the structure of the reconfigurable fabric.
Using low-radix OCSes as building blocks, \projname supports the forms of reconfiguration needed in training clusters including topology selection, workload adaptation, and failure resilience.
The cost of \projname scales with supported topologies and adaptations rather than with port count, breaking past the scalability barriers of current specialized ML networks.
We show through simulation that \projname-based deployments match the performance of fully provisioned packet-switched networks when training state-of-the-art LLMs at scale, while delivering significant cost savings using existing off-the-shelf OCSes, with strong bandwidth scaling and higher cost savings in the future.

\end{abstract}
\maketitle
\section{Introduction}

Training state-of-the-art AI models has become one of the dominant workloads in modern data centers, routinely involving tens of thousands of accelerators that exchange massive volumes of data at fine time granularity. To sustain this demand, practitioners increasingly rely on specialized interconnects rather than general-purpose networks, including High-Bandwidth Domains (HBDs), such as NVLink-based GPU clusters~\cite{nvlink} and Optical Circuit Switches (OCSes)~\cite{tpuv4}, providing high-capacity, reconfigurable connectivity at scale.

A defining characteristic of distributed training is its structured and phase-dependent nature. Modern training pipelines employ multiple dimensions of parallelism -- data, tensor, pipeline, and expert parallelism -- each inducing a distinct communication topology~\cite{skeletonhunter,99problems}. These are naturally time-multiplexed over the training iteration: from the algorithmic perspective, at any given phase, an accelerator participates in communication along a single dimension. Consequently, the network is not required to support all communication patterns simultaneously.

Existing interconnect designs, however, are built around the opposite assumption. Both HBD networks and current OCS-based systems provision a \emph{single monolithic fabric} that is sufficiently expressive to realize largely all-to-all connectivity at any time. In HBDs, this generality is provided by high-radix packet switches, whose aggregate bandwidth, power consumption, and cost scale poorly. In OCS-based designs, it is achieved through large ($N$\by{}$N$) optical switches that can dynamically reconfigure connectivity, but whose cost and control complexity grow quadratically with port count~\cite{mixnet}. The requirement to support arbitrary connectivity between the ports fundamentally limits their scalability.

This insistence on universal connectivity results in rigid and costly systems. Networks are provisioned to satisfy worst-case connectivity demands, even though most phases of training require only structured and relatively sparse communication. Moreover, existing solutions are typically engineered for fixed levels of fault resilience without offering the flexibility to trade availability for lower cost.

Direct-connect topologies, such as those used in Google’s TPU pods~\cite{tpuv4}, avoid monolithic switches altogether and achieve favorable cost and power characteristics. However, their fixed and limited connectivity constrains bandwidth efficiency and adaptability across different communication phases during training, highlighting the tension between expressiveness and efficiency in current designs.

In this paper, we evaluate the feasibility of an alternative fabric design which sidesteps the requirement for a universally expressive interconnect, achieving a data-center-scale fully optical network. Our fabric provides \emph{only the specific topologies required by training workloads}, instantiated at the times they are needed. We propose \textbf{\projname, \longprojname}. \projname is a unified data center-scale network architecture that is co-designed with the training process and constructed from commodity, ultra-low-radix optical circuit switches connected directly to the end-hosts. By reconfiguring these switches \emph{during the training iteration}, the network dynamically realizes a sequence of structured topologies, each suitable for a particular dimension of parallelism, and achieves large scale without incurring the quadratic cost of full $N$\by{}$N$ connectivity. 

This approach is motivated by three key observations. First, different parallelism dimensions are naturally time-multiplexed during training, allowing the traffic at each dimension to be handled independently of the others. Second, the reconfiguration latency of low-radix OCSes is sufficiently small to be amortized by the computation and communication phases of a training iteration, allowing intra-iteration reconfiguration with low performance cost. Third, low-radix optical switches are commodity, enabling construction of inexpensive optical networks today, without high-radix hardware. 

Recent work has converged on the insight that optical circuit switches offer compelling benefits for ML training systems, providing a future-proof network fabric with dramatically lower energy costs and cheaper bandwidth scaling than packet switches~\cite{mixnet,photonic-rails,topoopt,infinitehbd,scaling-silicon-photonic,tpuv4}. But all of them rely on full $N$\by{}$N$ fabrics or units with fixed topology. In contrast, we view the interconnect as an ensemble of structured topologies, each instantiated only when required by a specific parallelism dimension. Notably, \projname dedicates \emph{all the I/O bandwidth} of a network node to the active topology, departing from the common partitioning of scale-up vs. scale-out. As a result, it transitions from general-purpose connectivity to an explicitly time-multiplexed topology, defining a new point in the design space that prior systems did not explore.

Building large-scale systems from low-radix switches introduces fundamentally new design questions. First, what is the \emph{physical topology} to connect low-radix switches to enable \emph{logical topologies} that work well with collectives, while maintaining the number of chained OCSes under the maximum loss budget? Second, how to support flexible sizes of topologies to allow varying degrees of parallelism with different sizes of collectives, as well as to support multi-tenant fabric allocation? Third, how to maintain connectivity in the presence of processor, link, or switch failure? Finally, how to coordinate thousands of switches while avoiding  centralized bottlenecks and large failure domains.

In this work we first introduce several general concepts for building a training fabric from low-radix OCSes, and then demonstrate how they can be applied to build real deployments using off-the-shelf hardware. At a high level, each accelerator is connected to a static set of $k$ topologies suitable for different collectives, via a \emph{topology-selection 1\by{}$k$ OCS}. Each such OCS is reconfigured during training iterations to connect to a particular topology suitable for the running collective.  For example, we support rings, chains, tori and random expanders.  
Each such topology can be \emph{resized} to support multiple partitions via \emph{topology-adaptation 2\by{}2 OCSes} (\cref{sec:adaptation}). Importantly, when resizing one topology, we account for the complex interplay between it and the other topologies involving the same accelerator, while providing enough flexibility to support a variety of degrees of parallelism used in model training.  Finally, we demonstrate how to make topologies fault-tolerant via \emph{topology-resilience OCSes} (\cref{sec:resilience}). 

Topology adaptation and resilience OCSes require \emph{one-shot} reconfiguration -- they are reconfigured at job allocation time, and thus can be controlled  using a low-bandwidth centralized control plane. At the same time, the topology-selection OCS requires frequent reconfiguration. It is managed directly by the accelerator to which it is connected, \emph{without any global synchronization}, relying on the \emph{logical} alignment of the collective participants during computation (\cref{sec:control}).

Using these building blocks, we demonstrate a variety of possible deployments of training clusters at different scales and resilience configurations, starting from a small cluster with 16 GPUs, to medium-size 64/128/1024 GPUs, and up to 32K GPU data center-scale system. We conservatively estimate the cost of building them from off-the-shelf components, and evaluate via a large-scale simulation their performance on standard training configurations of six state-of-the-art LLMs.

\textbf{Our main findings} are as follows. \projname offers a wide variety of cost-effective options for small-scale systems, allowing one to pay only for the required capabilities and resiliency levels. Further, we show that even the most expensive configurations are cheaper than packet switch-based deployments by 27\% and 19\% for 4K and 32K-GPU systems. Moreover, \projname enables additional savings of up to 70\% for smaller scale clusters willing to compromise on flexibility and resiliency. We further show that significant cost savings over 70\% for clusters of varying scales can be achieved for future higher-bandwidth systems. 

This cost reduction comes with small or negligible performance overheads when training large-scale frontier models. Through large-scale simulation we evaluate the training performance of Qwen-2, Mixtral-7B and 22B MoE, Llama-8B and 70B, and LLama-4 Maverick, on medium and large scale \projname-based systems.  We show that \projname clusters consistently outperform static 3D torus topology emulating Google's TPU V4, performs on par with an ideal non-blocking packet switched network for all but Qwen-2. Qwen-2 is known to be particularly sensitive to the network performance~\cite{mixnet}. However even these overheads diminish with higher bandwidth interconnects. 

Our work clearly demonstrates that low-radix OCSes hold a great potential for building scalable optical networks. This work 
represents only the first step toward  the exploration of the design options in this space, and many challenges remain. Still, we 
believe that our approach paves the way toward practical low-radix systems in the near future.

\section{Motivation}
    
\subsection{Limitations of existing solutions}

\noindent\textbf{Packet-switched networks} form the conventional foundation of high-performance interconnects for distributed AI training, providing low latency and full bisection bandwidth through high-radix switching. However, sustaining these properties at scale has become increasingly expensive due to power density and packaging limits~\cite{lightmatter}.
Further scaling  depends on disruptive technologies~\cite{interconnect-survey}, such as wafer-scale integration~\cite{waferscale} or 3D-integrated photonics~\cite{lightmatter}, making continued growth in system size both economically and energetically challenging.

\noindent\textbf{Direct Connect Networks} connect processors directly without switches.
Such networks are currently used in Google’s TPU v4 clusters~\cite{tpuv4}. While attractive from a power and cost perspective, they impose a fundamental bandwidth penalty, called  \textit{bandwidth tax}. Achieving arbitrary connectivity requires multi-hop routing, which significantly reduces achievable throughput~\cite{cost-considerations-chip-network,topoopt,efficient-collectives}. Moreover, because training traffic varies across execution phases, any static direct-connect topology is inherently \emph{over-specialized}.

\noindent\textbf{Optical Circuit Switches} (OCSes) enable seamless bandwidth scaling and lower energy per bit, and have shown promise in training clusters~\cite{tpuv4,sip-ml,oddl,topoopt}. In some systems~\cite{tpuv4,topoopt}, reconfiguration is performed in a one-shot manner at the beginning of a training task in order to avoid the overheads of \emph{reconfiguration delay}, during which reconfigured portions of the network are temporarily unavailable. Recent works show that despite the reconfiguration delay,  OCSes can be used in training clusters, offering both cost efficiency in some scenarios today, but even more so opening the way toward bandwidth scaling in the future~\cite{mixnet,photonic-rails,topoopt,infinitehbd,scaling-silicon-photonic}.

However, most existing designs rely on monolithic high-radix switches, which incur costs that scale \emph{quadratically} with system size~\cite{computer-networks-textbook,cost-effective-ona,telescent}. At the same time, increasing the scale of OCS-connected networks is a challenge: optical loss constrains the number of switch stages that can be composed~\cite{optical-switching-papadimitriou}, while building high-radix switches leads to increased reconfiguration delay~\cite{scaling-silicon-photonic}.

\subsection{Opportunity: unique network requirements of ML training}
\label{sec:ml-comm-opportunities}

\paragraph{Non-overlapping collectives}
From an algorithmic perspective, collectives across different dimensions of parallelism -- data, tensor, pipeline, and expert -- are largely \emph{non-overlapping in time}: during any execution phase, accelerators typically participate in a single collective. In practice, some overlap can occur, e.g., when a GPU sends data to the next pipeline stage while also participating in a data-parallel \allreduce. Such overlap may improve performance because modern accelerators expose multiple network interfaces, such as NVLink and scale-out Ethernet, enabling concurrent use of independent fabrics.

However, when node bandwidth can be fully dedicated to each collective, overlapping communication introduces contention and can increase overall execution time. The only case where concurrent collectives may be beneficial is when a node only receives in one collective and only sends in another, which is uncommon. Most collectives used in training, including \alltoall, \reducescatter, and \allgather, involve both sending and receiving, making concurrent execution unlikely to yield benefits.

In OCS-based fabrics, this temporal separation across parallel dimensions creates opportunities to \emph{dynamically switch between collective-optimized topologies}.

\paragraph{Structured and repetitive communication patterns}
Communication in parallel ML training is highly structured. Each dimension of parallelism induces distinct collectives among fixed groups of participants. For example, Data-Parallel \allreduce involves only corresponding shards across Tensor-Parallel groups. These communication groups remain fixed across iterations and layers.

This property also holds for Expert Parallelism (EP), where communication occurs among fixed subsets of nodes within Expert Groups in systems such as Megatron-LM MoE and DeepSpeed MoE~\cite{megatron-moe,deepspeed-moe}. As a result, each parallel dimension can be served by a \emph{static topology optimized for its collective}.

\paragraph{Executing collectives over low-degree physical topologies}
Recent work shows that collectives can be efficiently supported using low-degree physical topologies~\cite{efficient-collectives,topoopt,tpuv4}. Many forms of parallelism are well supported by degree-two ring topologies, which enable bandwidth-optimal implementations of common collectives such as \allreduce, \allgather, and \reducescatter~\cite{ring-all-reduce}. Pipeline-parallel point-to-point transfers can be supported by linear topologies.

More complex patterns, such as \alltoallv in Mixture-of-Experts models, benefit from higher-degree topologies. Even then, low-degree graphs with small diameter can bound performance loss relative to fully connected systems~\cite{efficient-collectives}. For example, up to 57 nodes can be connected in a degree-8 graph with diameter 2, and over 1,100 nodes with diameter 4. As we discuss next, degree-8 topologies can be realized using commodity hardware.
Thus, \emph{sufficiently large communication groups can be realized without full inter-node connectivity}.

\paragraph{Intra-training OCS reconfiguration}
Prior work~\cite{photonic-rails,mixnet} shows that reconfiguration across collectives within a single training iteration incurs modest overheads, even with OCS reconfiguration delays on the order of 10~ms. Ongoing trends in model scaling further reduce the relative cost of reconfiguration, creating longer idle windows~\cite{photonic-rails} in which reconfiguration can be hidden. Our evaluation results (\Cref{sec:eval}) support this argument.

\paragraph{Summary}
Modern ML training workloads motivate \emph{specialized network connectivity}. Providing a limited set of low-degree topologies, rather than a fully connected $N$\by{}$N$ OCS fabric, avoids unnecessary cost while retaining required performance. We now show that such fabrics can be built efficiently using commodity hardware.

\subsection{Commodity hardware support}
\label{sec:hardware-trends}

\paragraph{Inexpensive low-radix OCSes}
Low-radix optical switches (e.g., 1$\times$2, 2$\times$2, and 1$\times$4) are commercially available at low cost. These devices also support fast reconfiguration: MEMS-based switches achieve sub-10~ms delays~\cite{switch-archive}, while newer technologies offer reconfiguration under 100~$\mu$s~\cite{glsun-magnetic-switch}. These trends make low-radix OCSes attractive building blocks for cost-effective optical fabrics.

\paragraph{Multi-port support in modern NICs}
Modern 800~Gbps Ethernet standards use multiple parallel lanes~\cite{800gbase-sr8}, enabling low-degree topologies. Commodity NICs from NVIDIA~\cite{connectx-8-port-split} and Broadcom~\cite{broadcom-port-breakout} support port splitting, allowing each lane to function as an independent 100~Gbps port. This enables moderate node degree without increased cost.

\paragraph{Summary}
Together, these trends suggest that specialized training networks built from low-radix optical circuit switches are both practical and advantageous, enabling scalable and bandwidth-efficient interconnects without the cost and complexity of high-radix designs.

\section{Design considerations}
   
Our design revisits several aspects embedded in packet-switched and high-radix optical fabrics. Below we summarize the key constraints and requirements that guide our design.

\paragraph{Unified vs. multiple fabrics}
Prior designs retain a high-bandwidth local fabric (e.g., NVLink) and add a separate optical fabric for global communication~\cite{photonic-rails,mixnet}. We instead target a unified fabric: splitting GPU bandwidth across multiple interconnects leads to fabric underutilization. As optical signaling is increasingly used even within local interconnects, the distinction between “local” and “global” networks is blurring, motivating a single reconfigurable fabric. 

\paragraph{Flexible collective sizes}
Training workloads induce collectives of varying sizes due to placement, scheduling, and resource availability. Tensor-parallel group size may be fixed by model constraints, while data-parallel collectives scale with allocated GPUs. Supporting multiple jobs on the same fabric similarly requires flexible partitioning. A topology optimized for 16 GPUs may be inefficient for 8, requiring \projname to support families of topologies parameterized by collective size.

\paragraph{Fault tolerance}
Direct-connect topologies are sensitive to component failures, which may break physical connectivity and render a topology unusable. This sensitivity, combined with the use of many low-radix OCSes, makes resilience to switch and NIC failures a key design requirement.

\paragraph{Insertion loss and fabric depth}
Insertion loss accumulates with each optical switch, limiting the number of OCS stages without regeneration. While long-reach powerful transceivers exist, they are costly and power-hungry. Consequently, deep multi-hop fabrics built from cascaded low-radix switches scale poorly and must be avoided.

\paragraph{Decentralized control plane}
Scalable management of many OCSes is required for correct behavior and fine-grain reconfiguration. Fully synchronous reconfiguration is difficult to achieve at data-center scale, motivating a decentralized control plane that does not rely on low-latency global synchronization.

\paragraph{Per-lane switching and effective degree}
OCSes switch individual fibers, each carrying one or more unidirectional lanes. Aggregating multiple lanes per fiber reduces switch count but increases transceiver cost and limits effective topology degree, constraining realizable topologies. Independent per-lane switching  is thus essential for building higher-degree topologies for \alltoall-type collectives.

\paragraph{Reliability as a cost tradeoff}
Existing interconnects provision high fault tolerance by default, increasing cost and enlarging fault domains even when applications could tolerate reduced availability. For example, NVLink-72 is used across both small deployments and large AI factories despite differing reliability needs. We therefore aim to provide flexibility in the reliability-vs-cost tradeoff.

\section{Design}

At a high level, \projname builds a network fabric to support varying communication patterns by dynamically \emph{selecting} a topology from a set of separate direct-connected topologies optimized for each collective, during a single training iteration. The set of topologies is defined at a deployment time, and cannot be easily changed. However, each topology can be \emph{adapted} to support different number of members, supporting different GPU allocations and mappings of parallel tasks to GPUs. To ensure \projname can remain functional after failures, each topology offers a \emph{resilience} mechanism that restores connectivity and enables to restart execution after remapping a subset of tasks to other GPUs \emph{with the same parallelism dimensions}. Both adaptation and resilience introduce a non-trivial interplay between different topologies, in order to ensure that the spatial relationships between different dimensions of parallelism are respected. Later, we show how these ideas combine in deployments of different scales (\Cref{sec:deployment}).

\subsection{Topology Selection}
\label{sec:selection}

\begin{figure*}[t]
\centering
\subfigure[Topology selection]{\includegraphics[width=.31\linewidth]{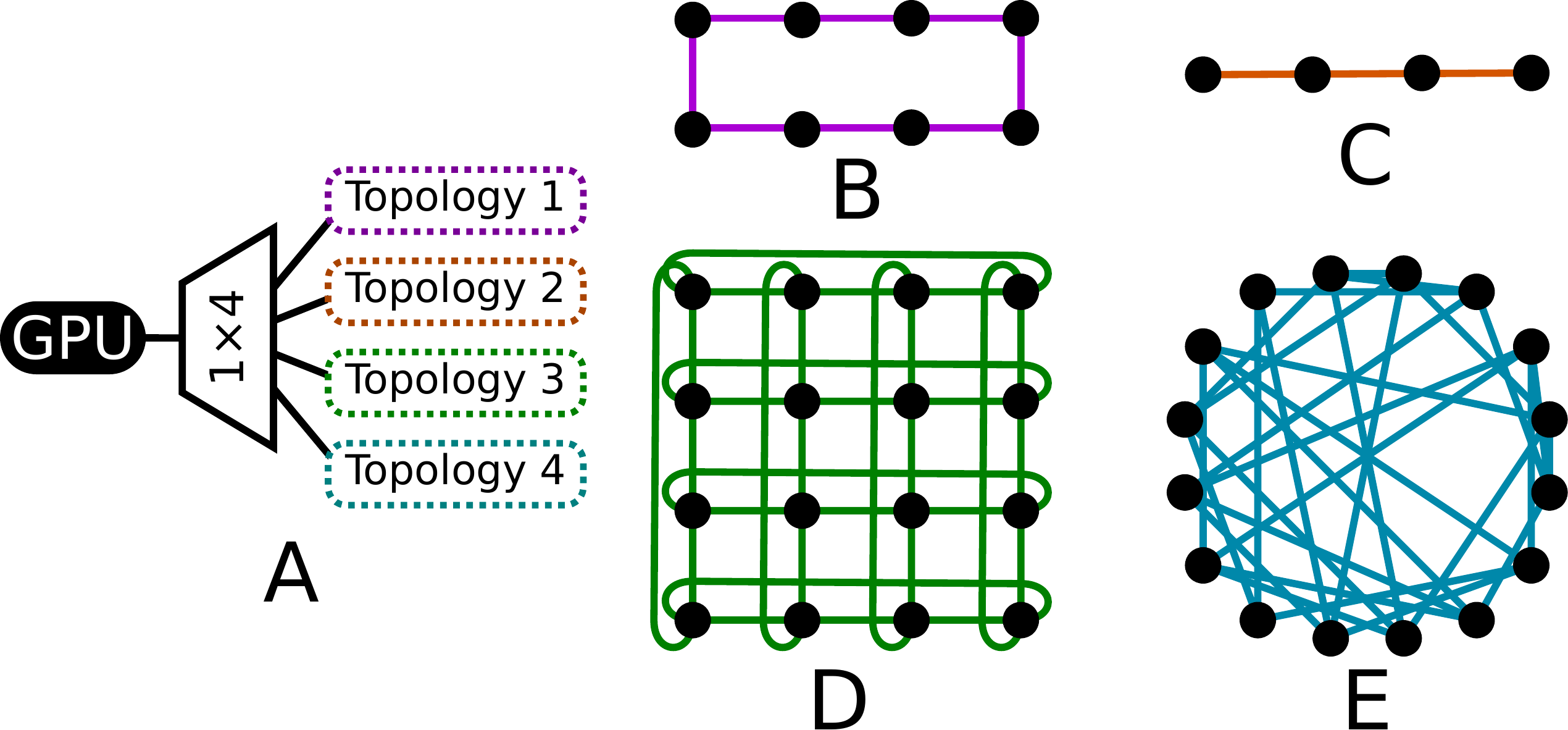}
\label{fig:topologies}}
\hspace{0.02\linewidth}
\subfigure[Topology adaptation]{\includegraphics[width=.31\linewidth]{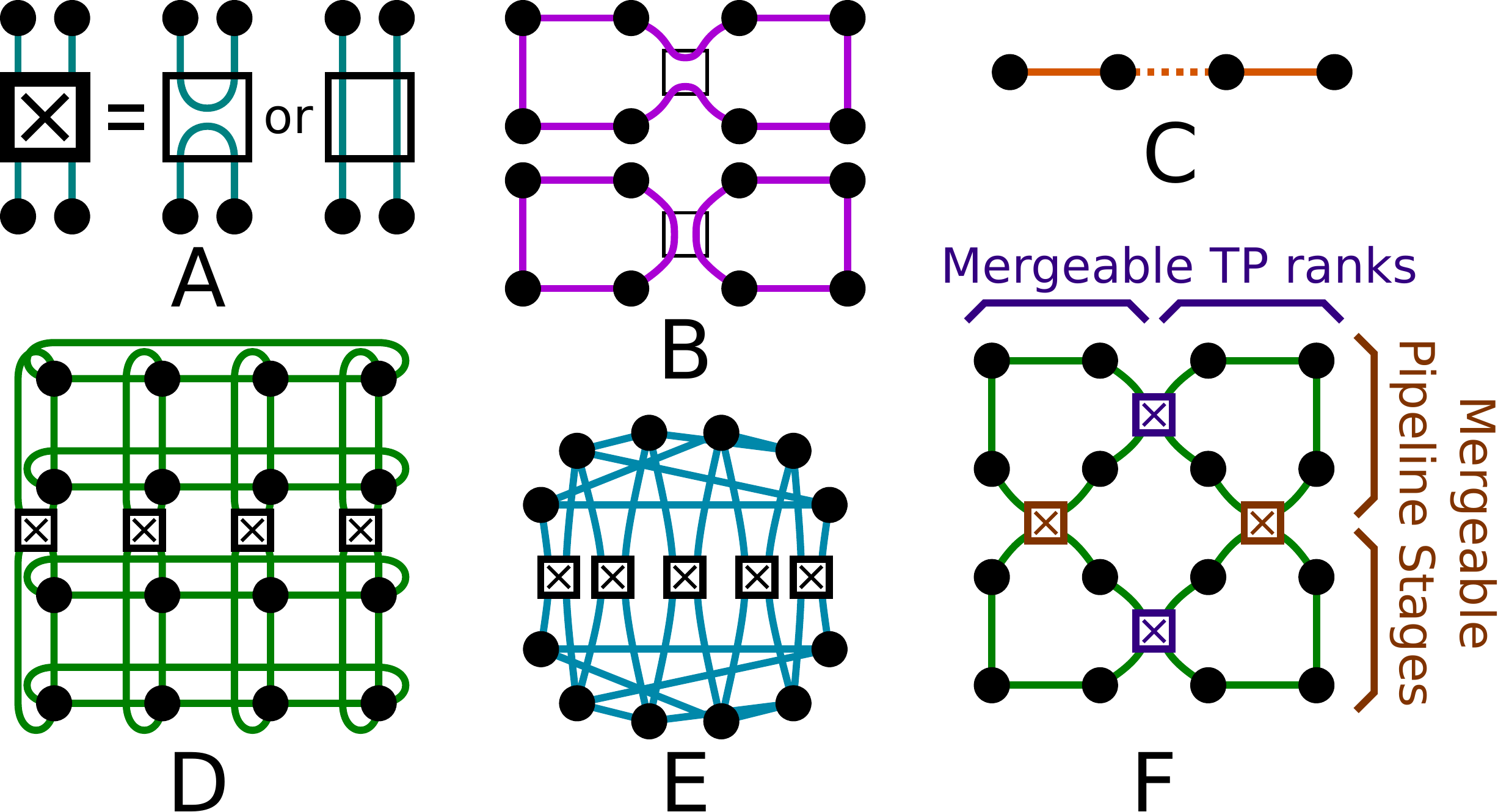}\label{fig:topology-adaptation}}
\hspace{0.02\linewidth}
\subfigure[Topology resiliency]{\includegraphics[width=.31\linewidth]{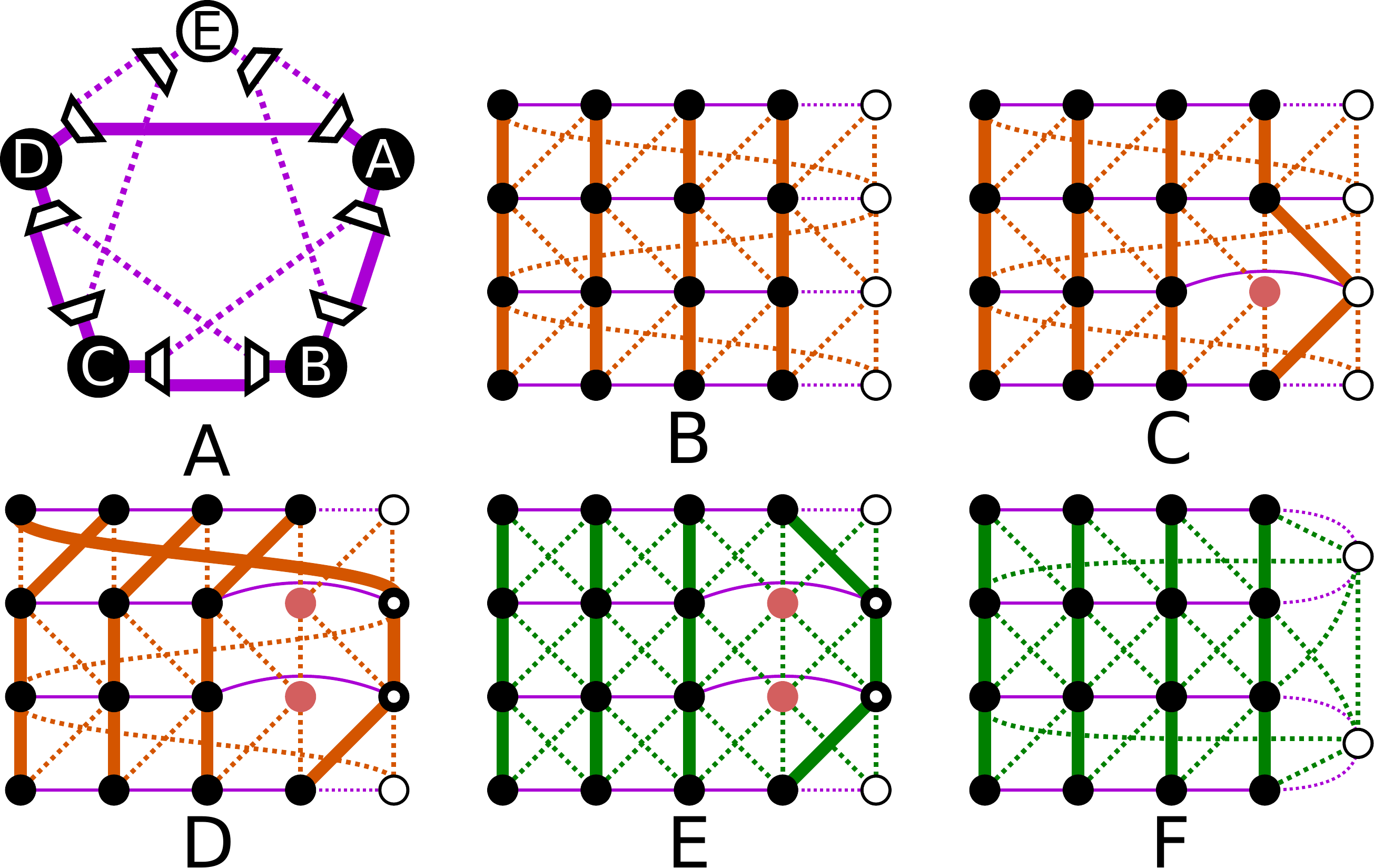}
\label{fig:topology-resilience}}
\vspace{-1em}
\caption{Main building blocks for \projname topologies}
\vspace{-1em}
\end{figure*}

We connect each fiber exiting the GPU NIC's optical transceivers directly to a 1\by$N$ topology selection OCS where $N$ is the number of optimized topologies. \Cref{fig:topologies} (A) shows an example where $N=4$. Each output of the 1\by$N$ OCS is connected to the corresponding topology, either directly to the 1\by$N$ OCS of another GPU or to other OCSes used for other forms of reconfiguration.

This design requires all the GPUs involved in a specific collective to switch to the respective topology to ensure connectivity. Thus,  before running the collective on a new topology, all 1\by$N$ OCSes attached to the participating GPUs must be configured to the corresponding topology. We discuss our approach to the OCS control plane in \cref{sec:control}.

One key consideration for \projname is the choice of optimized topology for each communication pattern. We choose topologies based on both the particular collectives run in each dimension of parallelism, as well as the number of available ports available at each  GPU, which defines the maximum \emph{degree} at each GPU in a topology. Note that degree differs from the radix of the 1\by$N$ OCS, which controls the number of topologies rather than degree. We examine several topologies, depicted in \cref{fig:topologies} and described below.

\paragraph{Ring and Torus}
Many popular collectives used in training can be implemented in a bandwidth-optimal fashion using a ring-based algorithm~\cite{mpich}. 
Rings only require a degree of 2 at each NIC, but can also be used when more lanes are available by constructing multiple \emph{parallel} rings to make use of all available bandwidth.

We apply ring-based topologies to dimensions of parallelism that use \allreduce, \allgather, and \reducescatter among small groups of GPUs, such as tensor parallelism with a degree of 8 or 16.
For dimensions with large groups of GPUs, we use a multi-dimensional torus topology. Using a Breadth-First Broadcast (BFB) schedule~\cite{efficient-collectives}, these collectives are bandwidth-optimal. This leverages higher per-GPU degree to minimize latency.

\paragraph{Linear}
In pipeline parallelism, linear demand patterns similar to a ring emerge, but they can be implemented on topologies that are not closed. This enables optimizations when considering adaptation and resilience.

\paragraph{Expander}
\alltoall and \alltoallv collectives used in Mixture-of-Experts models require different, low diameter topologies. Fully connected topologies are impractical for larger communicating groups. 
The best performance is achieved by using highly connected topologies with low average hop count, minimizing the \textit{bandwidth tax} suffered due to indirect routing. Prior work has found optimized topologies which achieve low hop count for a variety of group sizes and degrees~\cite{efficient-collectives}. However, the optimal topologies vary significantly with group size and degree. To avoid over-specializing on specific deployment characteristics in this work, we explore \textit{random expander graphs}. These graphs have low hop count with high probability~\cite{expander-diameter}.

\subsection{Topology Adaptation}
\label{sec:adaptation}

Different parallelization strategies induce different sizes of collectives. Consider a 16-GPU job with either two pipeline stages and eight data parallel replicas, or four pipeline stages with four data parallel replicas. The former runs optimally with linear and ring topologies of size 2 and 8 respectively, whereas the latter requires both topologies of size 4. 

We incorporate \textit{topology adaptation 2\by2 OCSes} (\cref{fig:topology-adaptation} (A)) to ensure that a single physical \projname network can be used to train jobs with varying parallelism degrees.
We use these OCSes to split and merge the topologies described in the previous section. 
We begin by describing how each topology can be resized in isolation, and then discuss special considerations arising from interactions between dimensions.

\paragraph{Ring and Torus} 
A single ring can be split into two rings half the size using a single 2\by2 OCS. This is done by routing two links of the ring through a 2\by2 switch, as shown in \cref{fig:topology-adaptation} (B). Reconfiguring the switch  either merges smaller rings or splits the large one.
The number of adaptation OCSes per GPU is equal to the number of parallel rings, i.e., fibers used by each link of the ring.
This approach can be used to split a large ring into multiple smaller sub-rings, by adding OCSes between different pairs of GPUs for each split. Only one 2\by2 switch is traversed along any given link, so the insertion loss from adaptation OCSes along reconfigured links does not depend on the total number of splits.

A torus topology is split simply by splitting each individual ring in one of the dimensions of the torus. The number of 2\by2 OCSes required is equal to the number of rings which must be cut, multiplied by the number of fibers per link. For example, if the 4\by4 torus shown in \Cref{fig:topology-adaptation} (D) uses 4 fibers per link, 16 2\by2 OCSes are required.

\paragraph{Linear}
A single linear topology can be split into two smaller linear topologies without the use of any OCSes. The link bridging the two sub-topologies is simply no longer used, as shown in \cref{fig:topology-adaptation} (C).

\paragraph{Expander}
We ensure that every link traversing between the two split halves of an expander is routed through a 2\by2 OCS, as shown in \cref{fig:topology-adaptation} (E). By reconfiguring these OCSes, we can convert two links which cross between the two halves into two new links, each of which is entirely located within one half of the topology. To do so, we introduce  \textit{splittable random expanders}, where exactly half of the links from each GPU traverse to the other half of the topology, but otherwise the links are selected randomly. These topologies perform on par or better than random expanders for \alltoall traffic (\Cref{sec:degraded-expander-eval}).
Expanders require the most 2\by2 OCSes for splitting -- a quarter of the total number of links in the topology, multiplied by the number of fibers per link.

\paragraph{Interactions between dimensions}
While topologies are multiplexed in time, their spatial distributions are linked. For example, under 3D parallelism, each GPU is assigned to a pipeline stage (PP) and a tensor-parallel (TP) \emph{rank}. Data parallel (DP) groups include all GPUs that share the same TP rank and PP stage. Reducing the TP degree reduces the number of ranks, requiring DP groups to be merged across what were previously different TP ranks.
Similarly, if the number of PP stages is reduced, DP groups that formerly corresponded to different pipeline stages must be merged.

While both cases involve merging DP groups, the particular groups being merged are different. This requires the DP topology to be adaptable in multiple independent ways, depending on whether the TP or PP degree is being changed. \Cref{fig:topology-adaptation} (E) shows one example of how this can be accomplished: 2\by2 switches are located at multiple positions on each DP ring, with one position used to merge DP groups corresponding to different TP ranks, and another used to merge DP groups corresponding to different pipeline stages. For torus topologies, the same approach can be used with multiple merge points at different positions on the same dimension, or by merging along different dimensions depending on which other form of parallelism is being modified.

\subsection{Topology Resilience}
\label{sec:resilience}

Failures may arise in different components impacting the system in different ways: GPU failures require tasks to be remapped to other GPUs, OCS failures may result in a degraded topology where some lanes are disabled, and NIC / transceiver failures combine both consequences.
In principle, each failure type could be handled differently. For simplicity, we first address GPU / NIC failures, and then show that the same resiliency techniques can also address switch failures by treating them as GPU failures.

We introduce redundant GPUs, as is already common in modern large-scale training designs~\cite{bloom-training,recycle-sosp24,oobleck-sosp23} which intentionally provision redundant hardware and spare capacity to tolerate failures. 
Our key challenge, however, is to recover the original topologies \emph{in all parallel dimensions simultaneously} after a failure, preserving their spatial relationships. This is essential to guarantee that a job can be restarted \emph{without modifying its parallel configuration}.

\paragraph{Resilient Rings}
To enable a ring topology to host a backup GPU, \projname uses \textit{resilient rings}.
Drawing on prior work~\cite{infinitehbd}, resilient rings include 1\by2 switches to allow the ring to skip over a single failed GPU, as shown in \cref{fig:topology-resilience} (A).

When one GPU in a resilient ring fails, the backup GPU is activated and compute tasks are reallocated among the remaining GPUs.
To preserve spatial relationships between dimensions, we choose to reallocate 
by shifting compute tasks over by a single GPU, and always shift in the same direction for a given ring. This ensures that a ring's rank of a particular task shifts by at most one GPU, which greatly eases handling three or more dimensions of parallelism.

\paragraph{Offsetting Links}
When a failure in a resilient ring causes the GPUs in the ring to shift tasks, orthogonal dimensions must also be reconfigured to account for the shift.
For example, consider a training system with a 2D organization (\cref{fig:topology-resilience} (B)) where  resilient rings are in  the horizontal dimension and the linear topologies are in the vertical dimension. If resilient rings are used alone, a GPU failure in a single ring would force all resilient rings to perform an identical shift to maintain an unchanged logical topology, while a second failure may make it impossible to restore the original logical topology even if it is located in a different ring.

To address this problem, we introduce diagonal \emph{offsetting links}, as shown in \cref{fig:topology-resilience} (B). In this example, a 1\by2 switch is added in front of each link in the linear (vertical) topology to create a single alternative, diagonal path. These diagonal paths are used to reroute the links from the linear topology, respecting the shifts created in the resilient rings. 
Offsetting links make each resilient ring an \emph{independent failure domain}. 

We consider two options. For a \emph{single} offsetting link, as in \cref{fig:topology-resilience} (B), the diagonal links are  constructed in alternating directions, and the shifts in the resilient rings are performed in alternating directions as well. This allows a single diagonal link to account for a shift in either resilient ring. However, some failures may sometimes make it impossible to connect corresponding ranks between both rows, as shown in \cref{fig:topology-resilience} (C). In this case, the dimension with offsetting links ends up shuffled to a different rank, which works for certain implementations of pipeline parallelism~\cite{megatron-lm}.

To overcome this, a 1\by3 switch is used to construct \emph{double} offsetting links, including diagonals in both directions as shown in \cref{fig:topology-resilience} (D). For double offsetting links, no combination of failures in adjacent resilient rings causes shuffling, ensuring spatial relationships can be fully recovered.

\paragraph{Shared Backups}
For smaller rings, maintaining an unused backup GPU for each ring may be too expensive.  
Instead, additional 1\by$N$ switches can be added  at the backup GPU, enabling it to be shared between $N$ different resilient rings to create a larger failure domain which can handle one failed GPU. \Cref{fig:topology-resilience} (E) shows an example in which backup GPUs are each shared between two rings.

\paragraph{Resilient Torus}
 Torus topologies can be made resilient using similar techniques: a single dimension of the torus can be converted into resilient rings, while the remaining dimensions are augmented with offsetting links. Although both types of links are now found within a single topology, the principles remain the same discussed above.

\paragraph{Resilient expanders}
So far resilience enabled the exact same logical topology to be maintained after failure. However, for expanders, this option is costly, and the degraded topology performs good enough.
 Thus, we include backup GPUs as part of the topology, and use them to route traffic indirectly even in the absence of failures. When a GPU fails, we shift tasks to alternative GPUs in a similar fashion to resilient rings, but without reconfiguring links. 
We evaluate  degraded expanders in~\cref{sec:degraded-expander-eval}.

\begin{figure}[t]
\centering
\includegraphics[width=\columnwidth]{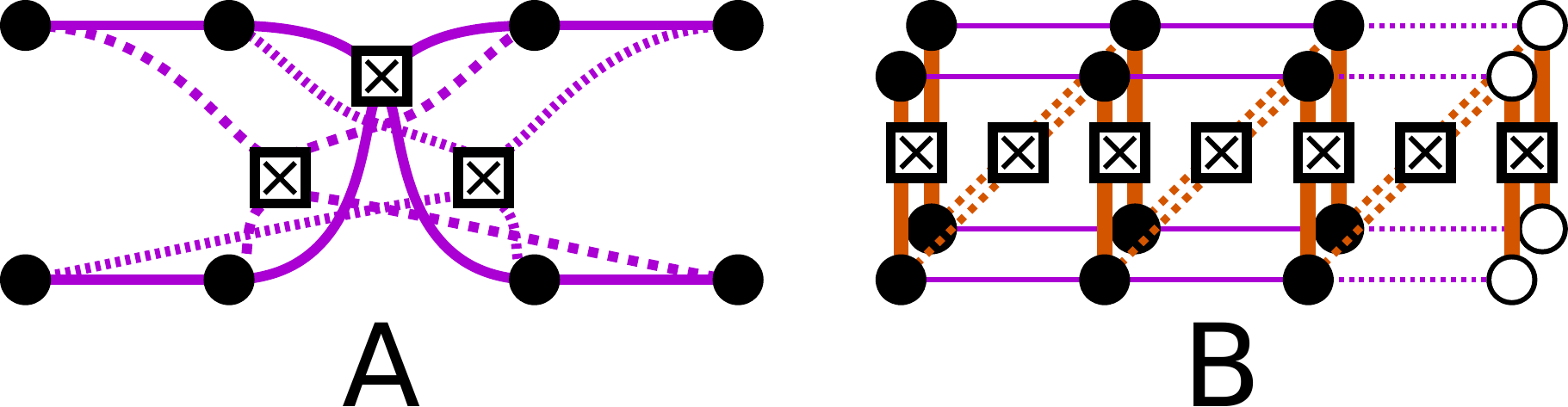}
\vspace{-1em}
\caption{(A) Merging resilient rings requires three sets of 2\by2 switches. (B) All 2\by2 switches on regular links are replicated on offsetting links.}
\label{fig:resilience-adaptation}
\end{figure}

\paragraph{Combining Resiliency and Adaptation}
Our resiliency strategies augment our base topologies with additional links. When performing topology adaptation, these additional links must also be reconfigured, requiring additional 2\by2 adaptation switches.
For example, merging a resilient ring requires three sets of 2\by2 adaptation switches to fully merge both regular and resiliency links, as shown in \cref{fig:resilience-adaptation} (A). In this case, the combined ring includes the backup GPUs from both rings, allowing multiple non-adjacent failures to be tolerated.

When using offsetting links, 2\by2 adaptation switches found on the regular links must be duplicated on the offsetting links, as shown in \cref{fig:resilience-adaptation} (B). 
For single offsetting links, this doubles the number of 2\by2 switches needed; for double offsetting links, the switches needed nearly triples. Network builders may balance the increased cost with the improved resiliency characteristics of double offsetting links when designing their network. Additionally, when merging or splitting the orthogonal resilient ring dimension, an extra set of 2\by2 switches is needed to reconfigure the offsetting links which cross the merge point between the rings. We show an example of this in \cref{fig:talos-72} (C).

\paragraph{Resiliency to Switch Failures}
Failures in 1\by$N$ switches can be addressed by treating them as a failure in the connected GPU. Meanwhile, 
because our resiliency strategies duplicate 2\by2 switches on both regular and resiliency links, a failed 2\by2 switch can be sidestepped similarly to when one of the GPUs on either end has failed, redirecting the logical topology to a link which traverses a different 2\by2 switch.

\paragraph{Failure Units}
Thus far we discussed failures of single GPUs. However, resiliency can also be achieved on the level of servers with multiple GPUs, or on the level of entire racks, e.g., in Google's TPU machine~\cite{tpuv4}. This can be accomplished by adding resiliency links only on links that traverse across different failure units. For example, by connecting \textit{corresponding} GPUs in different racks using resilient rings and offsetting links, an entire rack can be skipped at once, eliminating the need for redundancy within the rack. We also can combine multiple failure domains selectively. 

Using a larger failure unit reduces cost by eliminating the need for resiliency links within the unit. However, it also makes failures less isolated, since a single failed GPU in a unit makes the entire unit unusable. In practice, however, increasing the failure domain might be reasonable, since performing maintenance on a failed GPU often impacts multiple other GPUs anyway due to deployment constraints.

\subsection{OCS Control}
\label{sec:control}

\projname requires the coordinated control of tens of thousands of low-radix switches to build a unified network.
Since reconfigurations may be needed tens or even hundreds of times per second, centralized control is not a viable option.

To solve this challenge, we observe that different forms of reconfiguration have fundamentally different requirements. Specifically, the topology selection OCSes are the only ones that require frequent and fast reconfiguration, whereas the other two forms of reconfiguration are much less dynamic, and get invoked only at the job allocation, or when a failure occurs. Therefore, a single (slow) control plane is connected to all the adaptation and resiliency OCSes, but the topology selection OCSes are handled differently as follows.

\paragraph{Decentralized Control of Selection OCSes}
Topology selection is managed via a decentralized control mechanism delegated to the GPUs themselves. The 1\by$N$ topology selection switches are directly connected to the GPU-hosting servers, allowing the GPUs to actuate them autonomously. This architecture allows the interconnect to adapt immediately to application flow. Distributed control leverages the inherent synchronization of the application logic, which ensures GPUs participate in identical sequences of collectives at the same phases of training.

Reconfiguration occurs at the boundary between collectives. Once a collective operation finishes, the application determines if the subsequent collective should occur on a different topology. If so, the GPU immediately reconfigures its local 1\by$N$ topology selection switches. Before beginning the next collective, the GPU ensures that it receives a link-up event on all of its reconfigured links. Because 1\by$N$ switches do not send light down inactive links, a link-up event indicates that the neighboring GPU has also finished reconfiguring its topology-selection switch and is ready to begin the collective as well.  

Such an implicit synchronization for switching between topologies assumes that communications are performed via a collective communication library. By invoking a collective, the GPU explicitly introduces dependency among the other participants of the same collective. Thus it is aware exactly when the collective is over and it is safe to switch topologies.

\section{Deployment Scenarios}
\label{sec:deployment}

We now show how the principles of topology construction described in the previous section can be combined to produce a full network design with desired characteristics. We present several designs, ranging from a small cluster to a full datacenter-scale network.
Our designs are based on hardware available off-the-shelf today, including low-radix switches, transceivers, and NICs with support for splitting each lane into a separate logical port~\cite{connectx-8-port-split,eptc}.

\subsection{16-GPU Cluster}

\begin{figure}[tp]
\centering
\includegraphics[width=0.6\columnwidth]{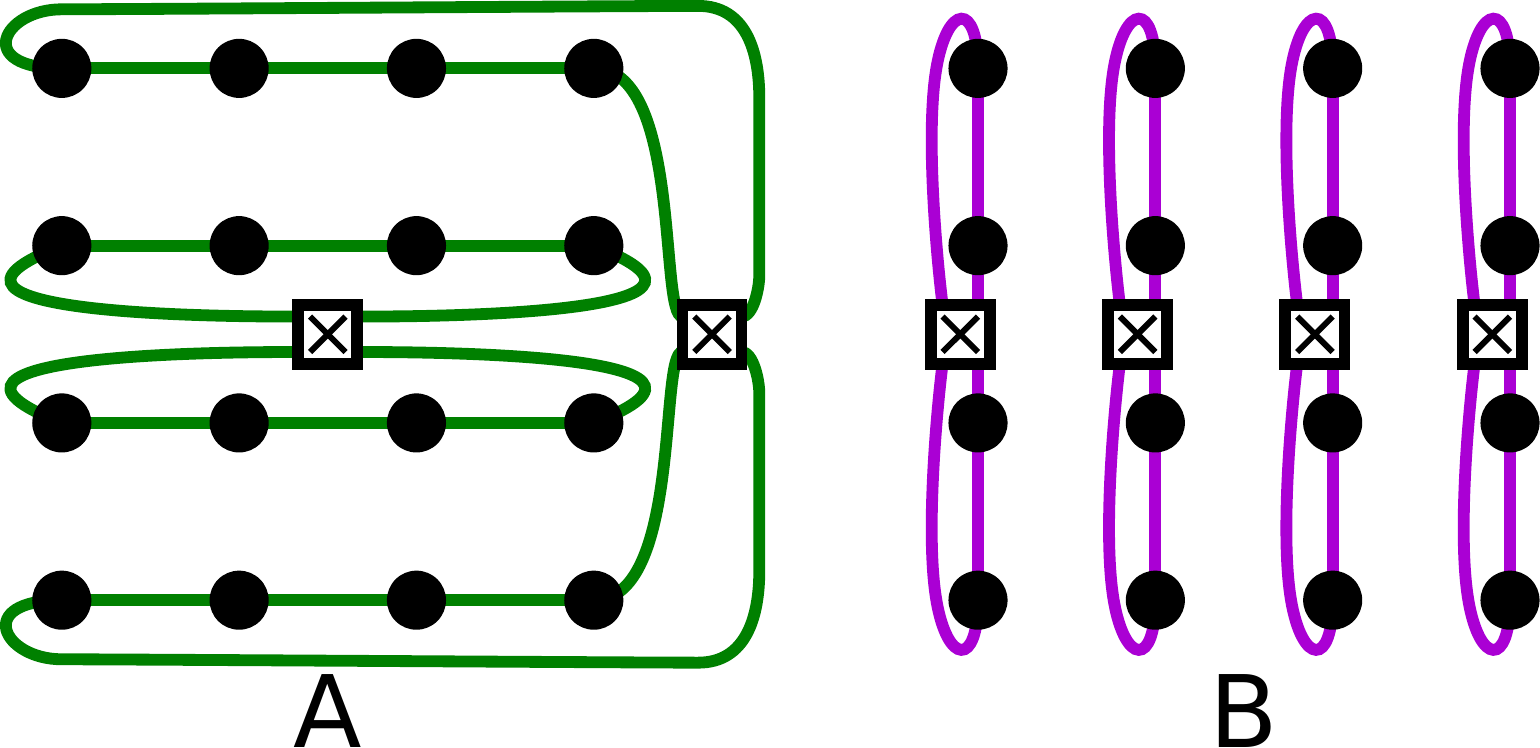}
\caption{16-GPU cluster with two orthogonal topologies for 2D parallelism. Horizontal (A) -- rings of sizes 4 and 8 GPUs, vertical (B) -- rings of size 4 or 2 GPUs.}
\label{fig:talos-16}
\end{figure}

We first consider the simplest deployment scenario with a 16-GPU cluster, shown in \cref{fig:talos-16}.
We implement two orthogonal resizable ring topologies of maximum sizes 4 and 8 respectively, each splittable into rings of 2 and 4. We assume the use of a transceiver which sends over 2 duplex fiber lanes, such as a 2FR4L~\cite{800g-2fr4}.
Each link in the topologies is composed of one 400Gbps lane comprising 2 fibers, one for each direction.
These topologies can be used to  train models using 2D parallelism, such as DP and TP with degrees 2,8; 4,4; or 8,2 respectively.

For such a small cluster, adding backup GPUs is likely unrealistic. Instead, there are options for reduced multi-GPU topologies even under GPU failures. For example, by configuring to use rings of size 4 in one dimension and rings of size 2 in another, the network is partitioned into two halves with 8 GPUs each. This maps well onto two NVIDIA HGX servers with 8 GPUs each, allowing one to be turned off for repair.

\paragraph{Cost}
We need  4 1\by2 switches to select between two topologies, one for each of the 4 fibers. Additionally, a total of 12 2\by2 switches are required,
an average of 0.75 per GPU. The total cost for these switches is \$125.50 per GPU. This is significantly below the cost of an 800 Gbps transceiver which would have been needed to connect to a packet switch, even without the cost of the switch itself. 

\subsection{Rack-scale cluster}

We now consider a cluster with  64 or 128 active GPUs, occupying 1-2 racks.
We add support for \emph{four dimensions of parallelism}: 
TP with degree 4, 8, or 16; PP with degree 1, 2, or 4 (or 8 for 128 GPU clusters), DP with degree 2, 4, 8, or 16; and expert parallelism (EP) with degree 8 or 16.
Note that the tensor and data parallelism-specific topologies are interchangeable.
To enable high-degree topologies for expert parallelism, we assume the use of a transceiver which sends over eight independent lanes, such as an FR8~\cite{800g-fr8}.

We show non-resilient and resilient configurations separately.  Recall  that a configuration is 
resilient if the same job can be instantiated without the changes in parallelism degrees after the failure. 
In a non-resilient system, certain topology sizes are impossible to instantiate at the original scale, even though smaller topologies would still work.

\begin{figure}
\centering
\includegraphics[width=\columnwidth]{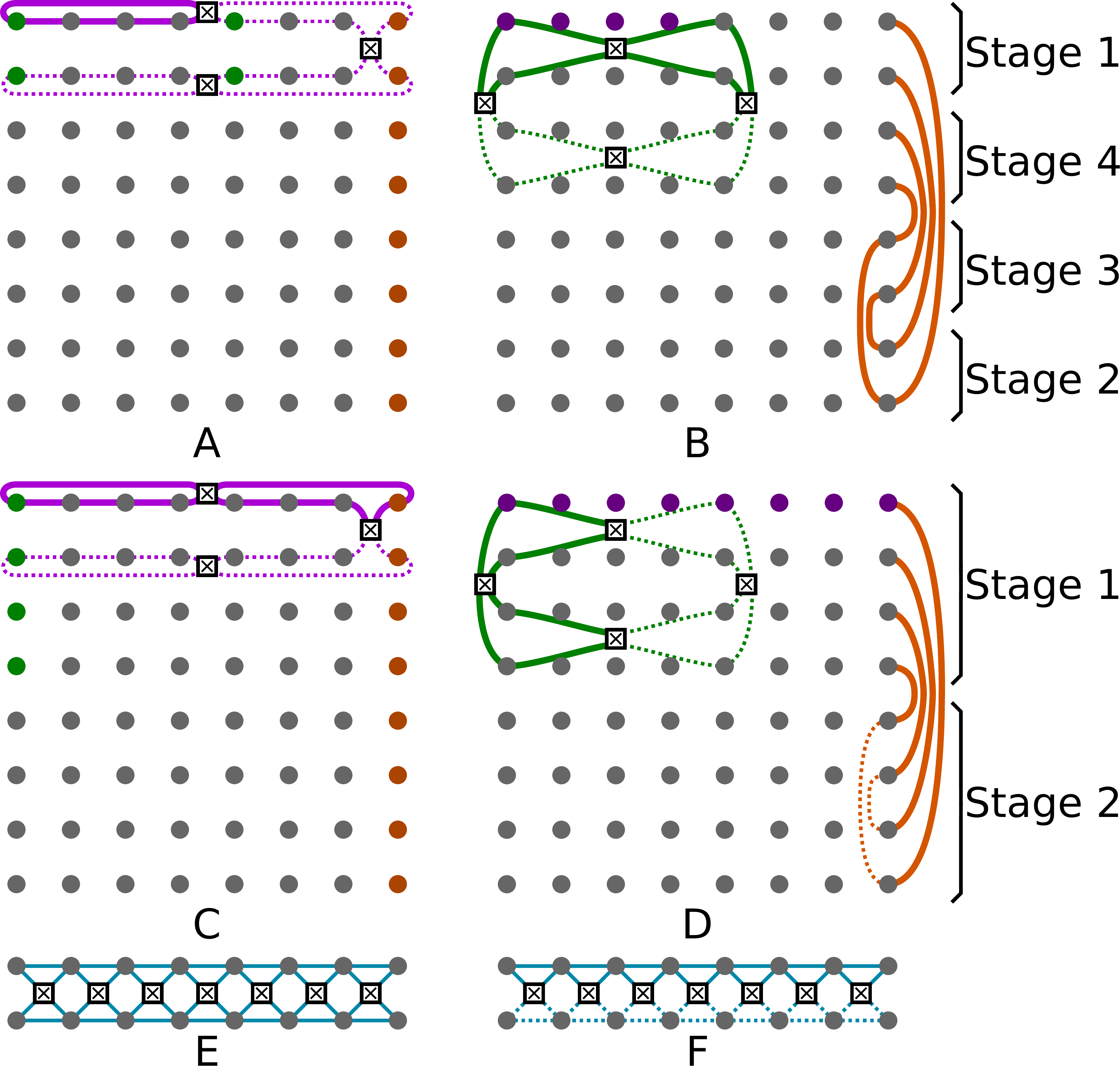}
\caption{Two configurations for a non-resilient rack-scale cluster: (A-B,E): TP=4, DP=4, PP=4, EP=16, and (C-D,F) TP=8, DP=4, PP=2, EP=8. 
Each row shows the same GPUs, with the TP (left) and DP+PP (right) topologies. Color indicates which nodes are connected by the topologies shown on the opposite side.
The EP topology is simplified in this figure.
}
\label{fig:talos-64}
\end{figure}

\paragraph{Non-resilient cluster}
To support four dimensions of parallelism, we use 1\by4 switches for topology selection, and implement a separate topology for each dimension as shown in \cref{fig:talos-64}. Tensor parallelism is supported via a splittable ring topology. Expert parallelism is supported via splittable random expander graphs, which can be split using 2\by2 switches into two complete graphs with 8 nodes. Pipeline parallelism is supported using a static linear topology, without OCSes.

Data parallelism is the most complicated dimension, as it must separately adapt to changes in both pipeline and tensor or expert degree.
To reduce the number of 2\by2 switches, we note that smaller degrees of pipeline parallelism leave some of the links in the linear topology unused. In some cases, these can be added to the data parallel topology to adapt to the change in pipeline parallel degree. As a result, some changes in PP degree are accomplished by reassigning links between topologies, without additional switches.

\begin{figure}
\centering

\includegraphics[width=\columnwidth]{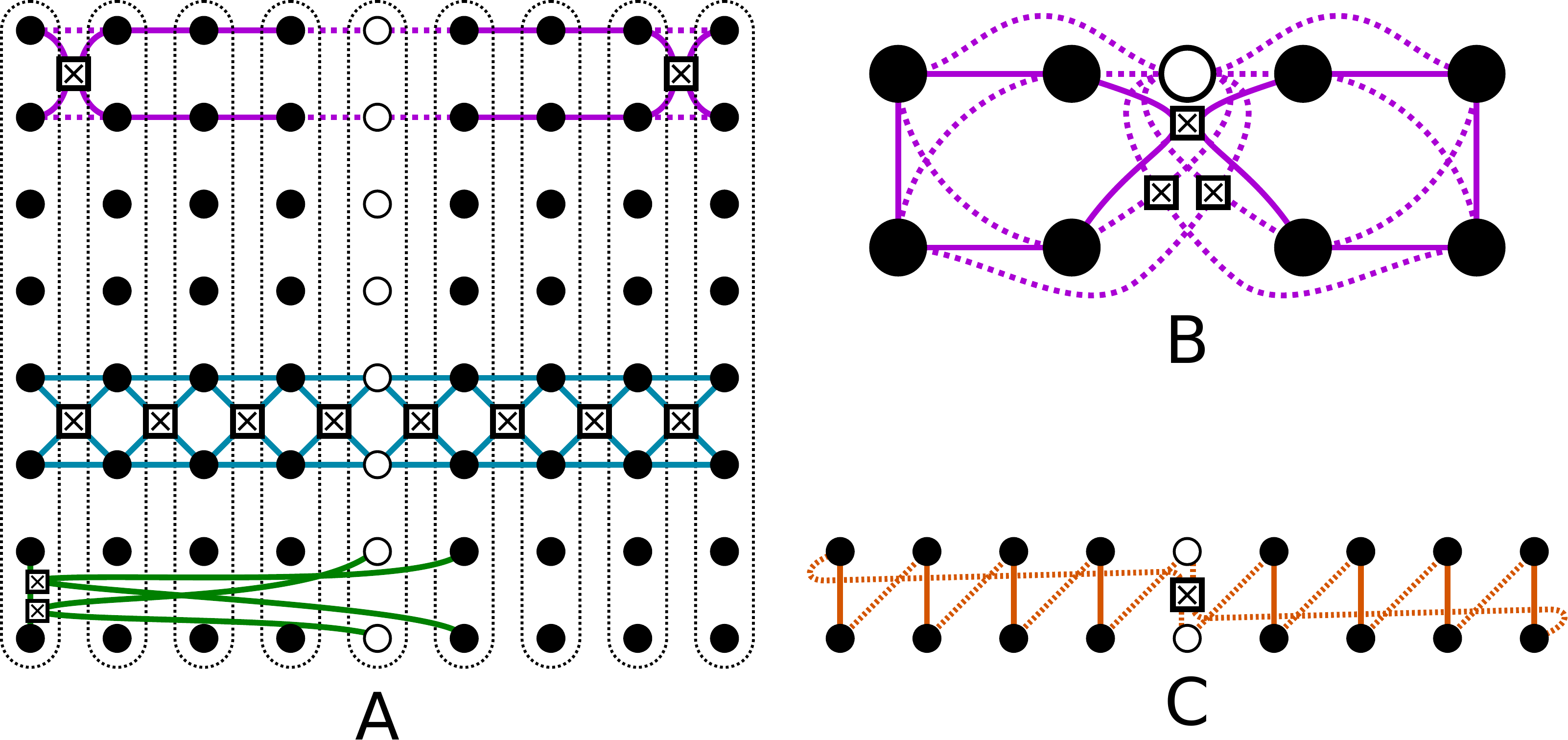}
\caption{Modifications needed to support node-level resiliency in the rack-scale cluster. (A) shows 9 nodes, each of which contains 8 GPUs. One node may fail in each rack. (B) shows a detailed representation of the resilient ring topology used for tensor parallelism. (C) depicts offsetting links used for cross-rack PP links.}
\label{fig:talos-72}
\end{figure}

\paragraph{Resilient cluster}
To build a cluster that is resilient to failures, we adapt our topologies to support \textit{node-level} resiliency. We assume that GPUs are housed in 8-GPU nodes, similarly to existing systems. 
To achieve this, we add an additional redundancy node to each rack, resulting in 64 active GPUs + 8 backup GPUs similarly to NVIDIA GB200 NVLink-72 systems. 
Our revised topologies are depicted in \cref{fig:talos-72}.

Our design locates overlapping tensor and expert parallel groups across all nodes in a rack. This enables expert parallel groups to be used as-is without reconfiguration, even after a node is taken offline, as discussed in \cref{sec:resilience}.

For tensor parallelism, we rely on resilient rings of size 8+1. To support a tensor parallel degree of 4, we modify these rings to allow them to be split into two rings of size 4 with a shared backup GPU, as shown in \cref{fig:talos-72} (B), which requires 3 2\by2 switches per link fiber.
When connecting two resilient rings together to support a TP degree of 16, we use two redundant sets of 2\by2 switches, ensuring that one set will be available even if one node fails.

For data parallelism, the topologies used for TP degree 4 span two nodes. To account for the possibility of a failed node, we double the number of 2\by2 switches used to transition between TP degree 4 and 8, allowing each node's DP groups to merge with those of two other nodes.

Finally, for 144-GPU clusters, each pipeline parallel group spans both racks. The links traversing between racks are no longer contained within a single node, meaning that they traverse between different failure units. We therefore add resiliency to these links through offsetting links, as shown in \cref{fig:talos-72} (C).

\subsection{Datacenter Scale}

Finally, we consider how \projname could be used to construct a large-scale network spanning many racks. We construct networks of this size using rack-scale units containing 64 active GPUs, as those described in the previous section. Our topologies support a TP degree of 4, 8, or 16; a PP degree of 4 or 8; and an EP degree of 16, 32, or 64. The DP dimension is determined according to the chosen configuration of the other dimensions and the total number of GPUs used.

\paragraph{Non-resilient cluster}
We make several modifications to the topology. First, we assign all GPUs within the same rack to the same PP stage. Second, we switch from ring topology for DP to using a 2D torus topology to reduce latency. At each GPU in the rack, half of the lanes are allocated to each dimension of the torus. We construct the rings for one dimension of the 2D torus entirely within a single rack, and use the other dimension to bridge between racks, creating a clean separation between reconfiguration due to changes in TP and PP degree. For especially large topologies, comprising tens of thousands of GPUs, we further move to a 
4D torus topology for DP, with three dimensions used to bridge between racks. Finally, anticipating that larger models will be trained on these topologies, we support higher EP degrees.

\paragraph{Node- and rack- resilient cluster}
The node resiliency offers the same level of resiliency per rack as the 72-node resilient topology. As long as the failures are not in the same rack, all the original topologies remain operational. However, if two nodes fail in a single rack, then the topology of the entire datacenter is compromised. Thus, we also support rack-level resiliency, allowing entire racks to be replaced with backups. Both forms of resiliency can be combined.

For node-level resiliency, we use internal rack topologies based on the 72-node resilient topology described in the previous section. We utilize offsetting links for all dimensions traversing across racks, including double offsetting links for the inter-rack dimension of the DP topology, to ensure that each rack constitutes an independent node-level failure domain. 
For inter-rack resiliency, we augment all inter-rack links with resiliency links produced at the rack level. For example, offsetting links connect to a corresponding GPU at a different rack, rather than a corresponding GPU on a different node.
We build the inter-rack data parallel dimension using resilient rings, and add offsetting links to the PP dimension (as well as remaining DP dimensions when using a 4D torus). 
When both forms of resiliency are combined, the number of offsetting and resiliency links needed on each inter-rack dimension must be multiplied to allow each link to be simultaneously shifted on the node and rack level.

For the cost analysis, we  choose to add one backup rack for every eight racks.
We specifically consider three sizes: 1024 GPUs across 16 racks (in this case using the PP dimension for resilient rings due to the small size of inter-rack DP rings); 4096 GPUs across 64 racks; and 32,768 GPUs across 512 racks using a 4D torus for DP.

\subsection{Cost Comparison}
\label{sec:cost-comparison}

We now evaluate the cost of our deployment scenarios, and compare them to several baselines. 
We do not include the cost of cables and NICs.
Our cost calculations are detailed in \cref{app:cost-calc}, and are based on publicly available prices for 800 Gbps Ethernet networking hardware, known costs for high-radix optical hardware~\cite{topoopt}, and quoted prices from OCS manufacturers. While 1.6 Tbps Ethernet is already available to large-scale operators, pricing is not public; we extrapolate the cost of higher line rates by proportionally increasing the cost of transceivers and the number of switches, following the bandwidth scaling achieved in 
multi-plane networks~\cite{nvlink}.

\begin{figure}
\centering
\subfigure[Cost scaling at 800 Gbps]{
\includegraphics[width=0.47\columnwidth]{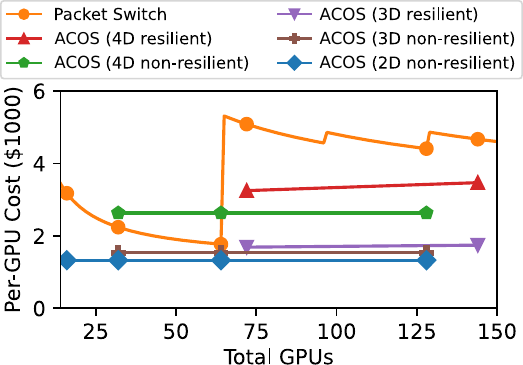}
}
\subfigure[\projname-16 relative to packet switch]{
\includegraphics[width=.47\columnwidth]{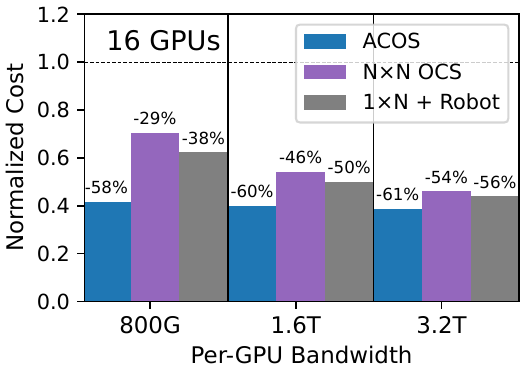}
}
\vspace{-1.2em}
\caption{\projname comparison for small scale systems.}
\vspace{-1em}
\label{fig:talos-16-cost-line}
\end{figure}

Our costs are normalized by the cost of an equivalent-bandwidth packet network. 
As shown in \cref{fig:talos-16-cost-line} (b), the cost per GPU of Ethernet networks at small scales is non-linear. This is because 800 Gbps Ethernet switches are available with 64 ports alone. Surprisingly, public pricing of 32-port switches is currently more expensive than 64-port switches~\cite{800g-ethernet-switch,32-port-800g-ethernet-switch}.

\paragraph{Small scale}
\Cref{fig:talos-16-cost-line} (b) shows the relative cost of the 2D-parallelism \projname setup for 16 nodes and two alternative optical baselines, normalized by the cost of a packet switch.
The first baseline is a single $N$\by$N$ OCS which can reconfigure in 50~ms~\cite{polatis}, representing traditional high-radix switching. 
Training will be worse than for \projname due to longer reconfiguration delay~\cite{photonic-rails}.
The second baseline attempts to combine the performance of low-radix switches during training with the flexibility of one-shot $N$\by$N$ reconfiguration. It uses 1\by2 switches to rapidly select between two topologies, each of which is configured using an $N$\by$N$ robotic patch panel~\cite{telescent-switch}, similarly to TopoOpt~\cite{topoopt}.
While cheap, it takes minutes to reconfigure. For a 16-node cluster, \projname is cheaper than both baselines, with nearly all of the cost arising from the transceiver at the end-host rather than the switches, and cheaper by more than half than respective packet switch.

\begin{figure}
\centering
\includegraphics[width=\columnwidth]{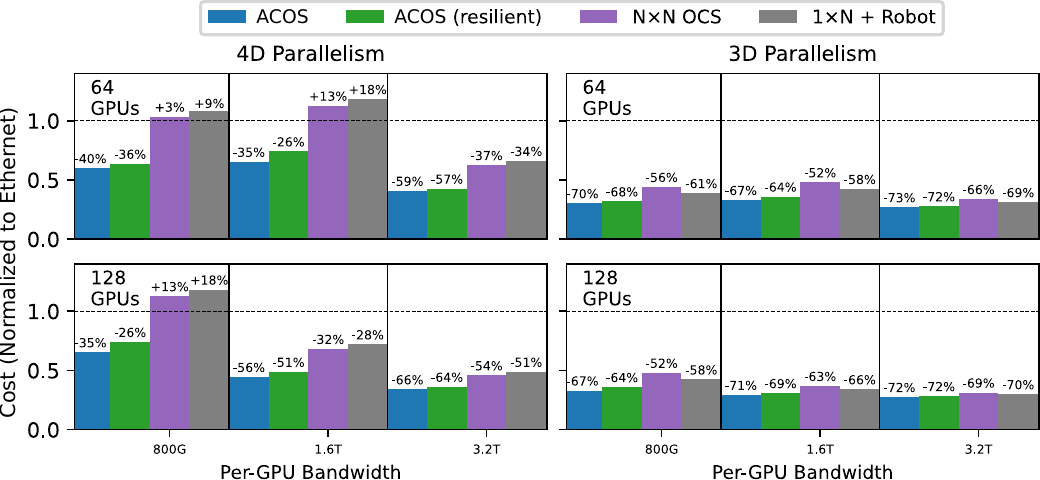}
\vspace{-2.2em}
\caption{Cost comparison for rack-scale deployments.
}
\vspace{-1em}
\label{fig:talos-64-cost}
\end{figure}

\paragraph{Rack scale}
\Cref{fig:talos-64-cost} shows cost comparisons for rack-scale clusters. As before, \projname is cheaper than both optical baselines. When implementing full 4D parallelism, many fibers must be switched per GPU, increasing cost for optical switching and increasing the gap between \projname and the cost of full $N$\by$N$ switching.
Additionally, the robotic baseline requires more patch panels to support more topologies.

\projname also outperforms the packet-switch baseline. For this cluster size, to ensure a fair comparison with \projname and $N$\by$N$ baselines, which support adding additional GPUs to ensure resiliency, we evaluate the cost of an Ethernet network which also includes the same number of resiliency nodes. As seen in \cref{fig:talos-16-cost-line} (a), the cost of an Ethernet network increases dramatically for topologies larger that 64 GPUs due to exceeding the port count of an individual switch. Instead, a two-layer topology is needed, requiring three times as many Ethernet switches per GPU and increasing cost beyond \projname.

If a deployment is not expected to be used for EP traffic, cost can further be reduced by using a 2-lane transceiver instead of an 8-lane one. This reduces the cost of optical switching technologies to far below that of a packet switch, with \projname costing less than a third of packet switches.

\begin{figure}
\centering
\includegraphics[width=\columnwidth]{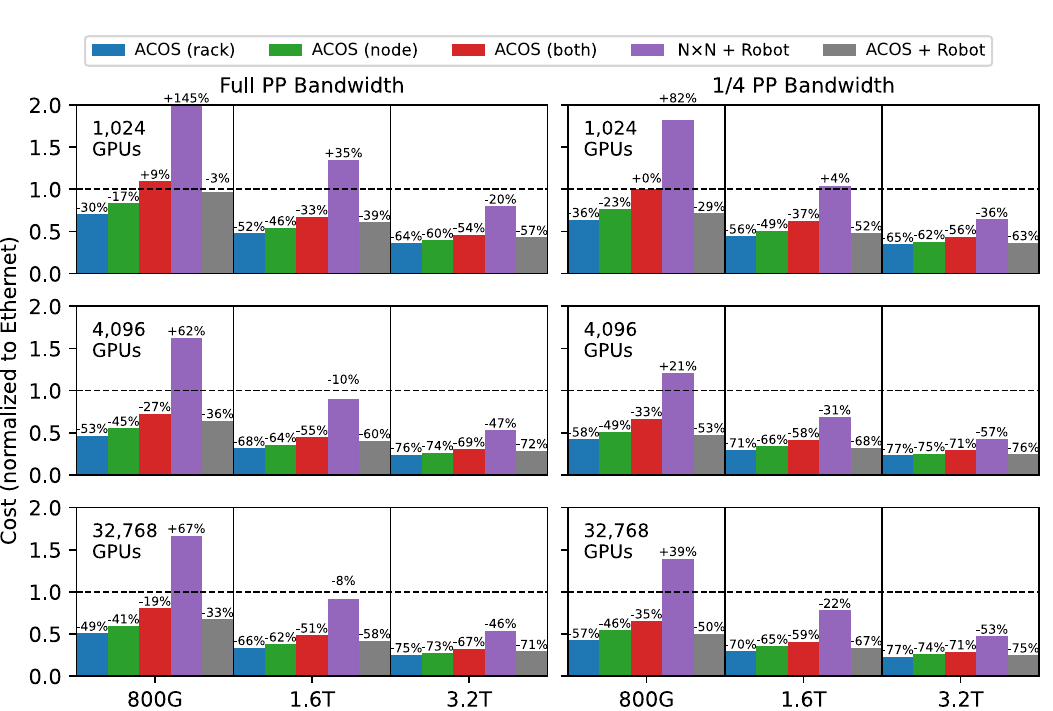}
\vspace{-2.2em}
\caption{Cost comparison for large-scale deployments. }
\label{fig:talos-dc-cost}
\vspace{-1em}
\end{figure}

\paragraph{Data center}
 For data centers full $N$\by$N$ circuit switching in a single stage is not practical. Instead, we evaluate two baselines: one in which each rack is internally connected by a 50~ms $N$\by$N$ circuit switch, with connections between racks controlled by extremely high port-count robotic patch panels, and one in which node-resilient \projname racks are connected to each other using robotic patch panels (reminiscent of Google's TPUv4 network architecture~\cite{tpuv4}, but with the ability to reconfigure the topology within the rack during training to maximize performance). 

As \Cref{fig:talos-dc-cost} shows, 
beginning at 4,096 GPUs, Ethernet must use a three-layer topology, greatly increasing the cost of Ethernet at this scale and giving \projname-based networks a clear cost advantage. \projname is also significantly cheaper than the $N$\by$N$ switch baseline, which requires a high port count on each rack's $N$\by$N$ switch to support both intra-rack and inter-rack connectivity. Finally, the combination of \projname with robotic patch panels performs quite favorably on cost. For settings in which several minutes of reconfiguration can be tolerated before each job, this combination offers an enticing combination of low capital cost, flexible inter-rack topologies, and per-collective optimized performance.

Another option to reduce the cost at this scale is to dedicate less bandwidth for the links over the pipeline dimension between the racks. This highlights the flexibility of our design to allow flexible I/O partitioning, without committing to the pre-defined scale-up/scale-out network bandwidth split.

\section{Evaluation}\label{sec:eval}

\paragraph{Simulation methodology}
We use Astra-SIM~\cite{astra-sim} to evaluate \projname across six state-of-the-art models. 
For this, we extended Astra-SIM to support reconfiguration to switch between different topologies used in \projname.  All presented results use the congestion-aware analytical backend, which we extended to support multi-dimensional topologies.

We generate traces for published standard training configurations for each model using MLSynth~\cite{mlsynth}. MLSynth generates  Astra-SIM compatible traces which match the I/O and compute operations based on the training configuration parameters,
which we specify in~\Cref{tab:model-configuration} in~\cref{app:eval}.

For MoE models, we use \alltoallv token distributions recorded during the training of Qwen-2 57B~\cite{qwen-57b} on a 64-node cluster to have realistic traffic skew that enables the evaluation of non-uniform MoE communication patterns. 

Topology selection is modeled as completely decentralized, without global synchronization. However, for simplicity, we introduce an artificial barrier that invokes the communication operation only after all GPUs in a given pipeline stage are configured to the correct topology. This offers a conservative estimate for \projname performance. The reconfiguration delay does not include link synchronization delay, following the methodology of prior works~\cite{mixnet,topoopt}.
We use a reconfiguration delay of \textbf{8ms}, matching the low-radix OCSes for which we received the pricing quotes in~\cref{app:cost-calc}.

The packet switch baseline refers to a non-blocking two level fat-tree topology. We balance the network load equally across all available paths. All topology comparisons are bandwidth equivalent, meaning that per-GPU bandwidth is split equally across all links according to the topology degree. 

We perform evaluations with existing 800Gbps networks that can be built at the time of this writing, and also evaluate higher per-GPU bandwidth of 1.6Tbps and 3.2Tbps, to assess the performance potential in the next generation systems as well as existing specialized HBD networks. We model GPU computing capacity according to NVIDIA's H200 GPU~\cite{h200}.

\subsection{End to End Benchmarks}
\begin{figure*}
    \centering
    \includegraphics[width=\linewidth]{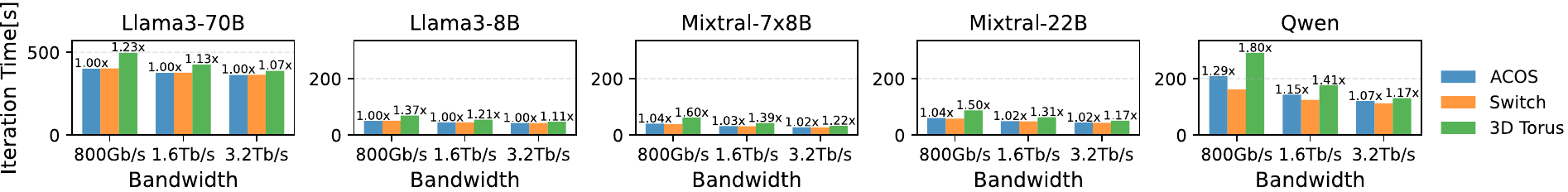}
    \vspace{-1.5em}
    \caption{Performance on a 64-GPU cluster. Bars are labeled with slowdown relative to switch.}
    \label{fig:e2e-results}
\end{figure*}

\paragraph{64-GPU cluster}
We first evaluate five medium-sized models on a cluster with 64 GPUs, following the design described in \cref{sec:deployment}. 
\Cref{fig:e2e-results} presents the iteration time for every model. 
\projname has no overheads when running the dense models (Llama3-8B and 70B) which require only 3D parallelism and do not use the expander topology. \projname constructs exactly the rings in which the 3D parallel training  communications are performed, while the structure of the training allows hiding the reconfiguration time entirely.

Note that the torus-based topology is consistently slower. Bandwidth is statically split across dimensions, so only a third of the bandwidth is available in each collective.
This problem cannot be easily fixed because the traffic flows across the same torus dimension, and spraying across the other unused lanes  to communicate between the nodes will ultimately require sending traffic over the dimension being actively used thus resulting in congestion. 

MoE models Mixtral~\cite{mixtral} and Qwen-2~\cite{qwen-57b} show higher overheads for both \projname and the 3D torus.
We attribute these overheads to the \alltoallv collective used in MoE layers. The bandwidth tax of routing large collectives through the expander graph causes significant overheads. We analyze the \alltoallv collective on an expander separately below.

The overheads are higher in Qwen-2 than in Mixtral, because of the much lower compute/I/O ratio in Qwen, and because it accesses 8 experts from 16 GPUs over an expander topology, whereas Mixtral accesses 2 experts over 8 GPUs. Coincidentally, when the 16-GPU expander is split in half, 2 sets of fully-connected GPUs get created. Therefore,  Mixtral does not suffer from the overheads of the expander topology. 
Using higher per-node bandwidth reduces the communication overheads for Qwen down to 7\%. 

\begin{figure}[t]
    \centering
     \subfigure[Performance comparison]{
        \includegraphics[width=.53\linewidth]{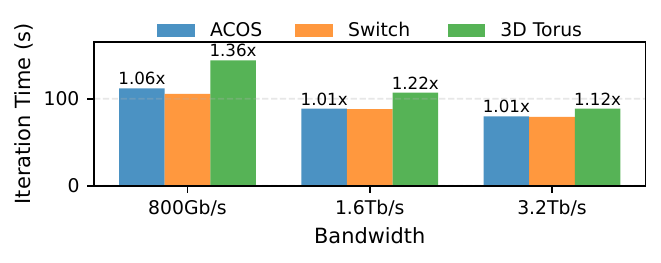}
    }%
    \subfigure[Execution breakdown]{\includegraphics[width=.42\linewidth]{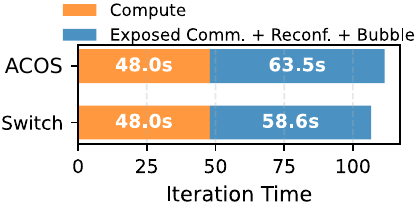}}
    \vspace{-1em}
    \caption{Llama 4 Maverick model on 1024 GPUs.}
    \label{fig:llama4}
\end{figure}

\paragraph{1024-GPU cluster}  \Cref{fig:llama4} compares the training iteration of the Llama-4 Maverick~\cite{llama4} frontier model trained on 1024 GPUs. This is a mixed MoE-dense model which  
uses expert parallelism only on half of its layers. Overhead is relatively low, disappearing entirely at higher per-GPU bandwidth.

\paragraph{MoE Performance Analysis} 
We compare Qwen-2 performance under three scenarios: original MoE traffic pattern over 16-node random expander, uniform traffic with the same total per-collective bandwidth over 16-node expander, and the original MoE traffic pattern but  with all 16 nodes forming a fully-connected topology. We observe that the skewness of the MoE traffic distribution has a minor contribution to the observed performance slowdown.
The problem is caused by multi-hop routing in the topology. A  fully connected topology shows a $17.7\%$ speedup compared to \projname, and effectively becomes as fast as the packet switch (not shown). Optimizing MoE over expanders poses a challenge we leave to future work. 
We further observe that increasing sequence length from 4K to 16K improves Qwen's performance.
The exact measurements are in \cref{sec:appendix:moe}.

\subsection{\alltoall collectives over expanders}
We  now study the performance of \alltoall over expanders used in \projname. We simulate an \alltoallv collective using the recorded MoE traffic distribution across three topology sizes. 
\begin{figure}
    \centering
        \includegraphics[width=.9\linewidth]{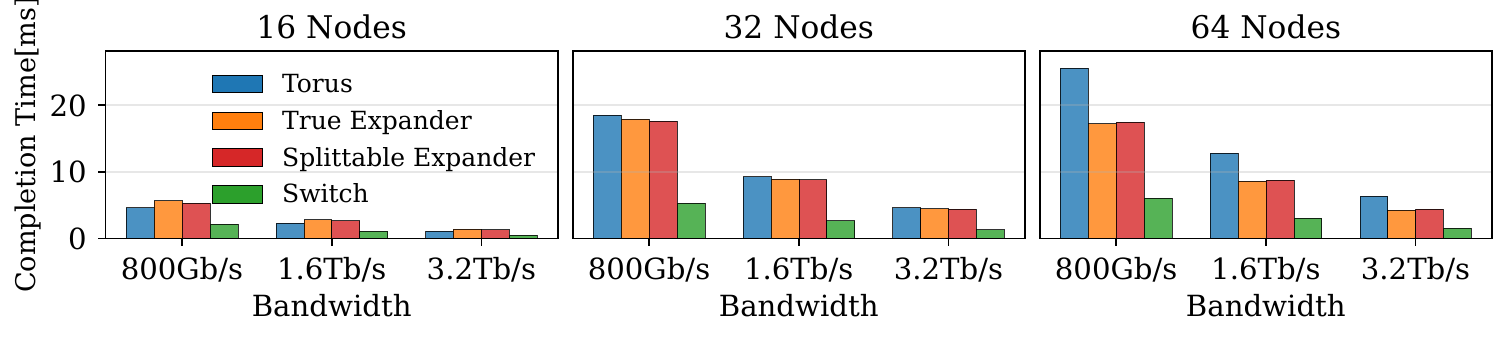}
    \vspace{-1em}
    \caption{\alltoallv performance vs. expander sizes.}
    \label{fig:all2all-ubench-analytical}
\end{figure}
\Cref{fig:all2all-ubench-analytical} shows that splittable expanders used in \projname perform nearly identically to true random ones in terms of the \alltoallv performance. While expanders are slower than the switch, they fare well compared to a 3D torus due to the higher diameter of the torus. The gap with switched topologies narrows with increased bandwidth.

\begin{figure}
    \centering
    \includegraphics[width=.8\columnwidth]{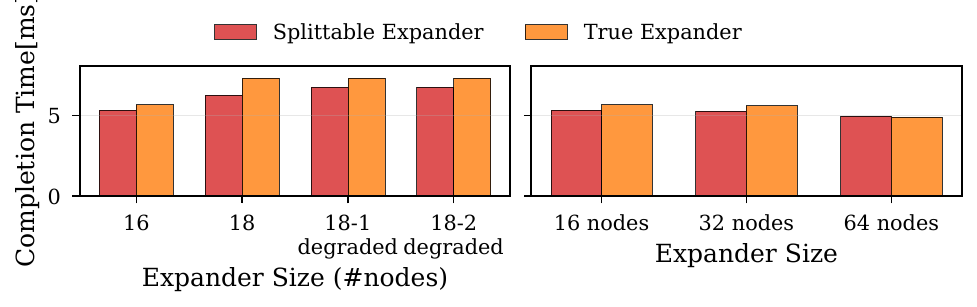}
    \vspace{-1em}
    \caption{16-GPU \alltoall on different topologies.}
    \label{fig:smaller-num-ranks-on-64-topos}
\end{figure}\label{sec:degraded-expander-eval}

In the GPU or node level resilient configurations of \projname,  not all nodes of an expander are involved in the collective, but all participate in routing. \Cref{fig:smaller-num-ranks-on-64-topos} shows the effects of running \alltoallv collectives among only a subset of the topology. \Cref{fig:smaller-num-ranks-on-64-topos} on the left shows that a degraded version of the GPU level resilient expander topologies of 18 GPUs, with one or two GPUs failed and unable to forward traffic, increases the completion time of the collective by $8\%$ and $7\%$, respectively. 

As the expander size is fixed during deployment, it may be necessary to run collectives between a smaller number of nodes on large topologies. \Cref{fig:smaller-num-ranks-on-64-topos} shows that running a 16-node \alltoallv on over-sized expanders results in similar or improved performance. The improvement is caused by lower chance of congestion and higher path diversity in larger topologies.  Note that this advantage would disappear if multiple \alltoallv collectives were invoked at the
same time over different nodes, and they would also contend for bandwidth (not shown). Therefore expander splitting is still necessary to ensure performance isolation among collectives. 

\section{Discussion and open questions}

\projname deployments are designed to support common training configurations. By specializing in this manner, \projname achieves a  combination of high performance and low cost for all-optical technologies.
Admittedly, such a specialization reduces flexibility of parallelization configurations. 
At the same time, the particular deployment scenarios in this section can be adjusted to include more or fewer switches for greater generality or lower cost. 

Our cost comparisons are made assuming current off-the-shelf hardware, but they can be reduced significantly with customized hardware. For example, Google already makes use of optical circulators~\cite{tpuv4,lightwave-fabrics}, halving the cost of optical switching. Broader interest could result in commercial availability of such technologies.

Resiliency is an important issue. We believe that the simplicity of low-radix OCSes may allow extremely high per-component reliability, compensating for the massive quantity of such OCSes we use. At the same time, if 0.1\% of GPUs in a cluster are faulty, clusters using both rack-level and node-level resiliency offer high availability. For example, a cluster with 1,024 active GPUs can create a pristine logical topology with over 99.9\% probability, while even a 32,768-GPU cluster can create a pristine topology with 98.9\% probability (\cref{app:resiliency}). However, if these failure rates are prohibitive for the pure low-radix OCSes-based designs, the use of $N$\by$N$ switches for inter-rack resiliency may be more practical.

Switch faults render the links that traverse them unusable. Since each link traverses only a small number of switches to limit optical loss, localizing switch failures is not different from localizing standard link faults.  Correlating faults across nodes can further assist in localizing failed switches.

Finally, our designs and evaluation are theoretical, focusing on developing the principles of \projname.
This leaves several practical considerations to future work.
In particular, augmenting collective communication libraries to support  reconfiguration is also left to future work.
Similarly,  while our design for distributed reconfiguration is possible in principle, it depends on hardware support to rapidly detect and report link-up events to the application which we do not evaluate. 
Microsecond-scale transceiver lock following reconfiguration has been demonstrated~\cite{rotornet}, but not in new transceivers hardware. We currently ignore this engineering challenge similar to prior work~\cite{mixnet,topoopt}.

\section{Related Work}

Several designs for optical networks for machine learning, including \textbf{TopoOpt}~\cite{topoopt} and SiP-ML's \textbf{SiP-OCS} design~\cite{sip-ml}, connect nodes to an $N$\by{}$N$ circuit-switched fabric and configure this fabric once at the beginning of each training task. Similarly, Google's \textbf{TPU v4} network~\cite{tpuv4} uses high-degree 3D MEMS switches to build variations of 3D torus topologies which are static during training.

Other designs cater to specific collective topologies, i.e., SiP-ML~\cite{sip-ml} and InfiniteHBD~\cite{infinitehbd} are both designed to  implement rings. 
However, this rigid design limits them from supporting non-ring workloads such as EP.

\paragraph{Intra-iteration reconfiguration}
\projname shares many of its motivations with \textbf{Photonic Rails}~\cite{photonic-rails} which aims to support ML collectives through topology-specific topologies by observing that intra-training reconfiguration is affordable. 
However, unlike \projname, Photonic Rails acts as a rail-optimized fully-connective network. 

MixNet~\cite{mixnet} combines an $N$\by$N$ OCS with packet-switched HBDs to efficiently support EP traffic.  It reconfigures during the training process based on the predicted traffic matrix during a given layer. MixNet focuses on EP, but depends on existing HBD and scale-out networks to both fill gaps in the OCS topology and support other collectives.

\paragraph{Optimized algorithms and topologies for collective communication}
Optimizing collective communication algorithms for direct-connect topologies is a long-standing research topic~\cite{reccl,tacos,taccl,efficient-all-to-all}. Further, \cite{efficient-collectives} co-designs the topology and communication algorithm to achieve optimal performance. \projname motivates expanding on this work to account for ease of topology adaptation and fault-tolerance.

\section{Conclusion}
We show that data center–scale training networks can be built from  arrays of low-radix optical switches while achieving competitive performance and cost relative to existing packet switches and high-radix OCS designs. 
Our results demonstrate that relinquishing full connectivity enables a scalable and efficient interconnect for modern AI training.

\noindent\textbf{Ethics:} This work does not raise ethical issues.

\bibliographystyle{ACM-Reference-Format}
\bibliography{bibliography}

\appendix
\crefalias{section}{appendix}

\section{Cost calculation}
\label{app:cost-calc}

\begin{table}
\begin{tabular}{|l|r|}
\hline
Item                               & Cost     \\ \hline
SR8 transceiver (100m)~\cite{800g-sr8}           & \$650    \\ \hline
DR8 transceiver (500m)~\cite{800g-dr8}          & \$850    \\ \hline
FR8D transceiver (2km, 8 lanes)~\cite{800g-fr8}           & \$1,100  \\ \hline
2FR4L transceiver (2km, 2 lanes)~\cite{800g-2fr4} & \$1,200  \\ \hline
64-port Switch~\cite{800g-ethernet-switch}        & \$30,000 \\ \hline
\end{tabular}
\caption{Costs for 800 Gbps Ethernet Equipment}
\label{tab:ethernet-cost}
\end{table}

This appendix presents the data used to generate our cost comparisons. First, \cref{tab:ethernet-cost} provides a list of publicly available costs for 800 Gbps Ethernet hardware which we use for our cost calculations. For Ethernet networks, we assume short-range 100m transceivers are used for links to leaf switches, while links to spines and super-spines use longer-range 500m transceivers. For circuit-switched networks, we use longer-range transceivers to obtain a higher optical loss budget. Depending on the number of lanes used in the topology, we select either the FR8D transceiver or the 2FR4L transceiver.

\begin{table}
\begin{tabular}{|l|r|}
\hline
Item                               & Cost     \\ \hline
$N$\by$N$ OCS~\cite{polatis,topoopt} & \$520 per duplex lane    \\ \hline
Robotic patch panel~\cite{telescent-switch,topoopt} & \$100 per duplex lane    \\ \hline
1\by2 switch & \$22 per switch \\ \hline
1\by3 switch & \$68 per switch \\ \hline
1\by4 switch & \$70 per switch \\ \hline
2\by2 switch & \$50 per switch \\ \hline
\end{tabular}
\caption{Costs for optical circuit-switching equipment.}
\label{tab:ocs-cost}
\end{table}

Next, \cref{tab:ocs-cost} provides the costs we use for optical circuit switches. In the case of the $N$\by$N$ switch and the robotic patch panel, costs are per duplex lane (each consisting of two fibers). On the other hand, the costs for low-radix switches are for individual switches, each of which can switch a single fiber. For low-radix switches, our costs directly reflect quotes we have received from switch manufacturers for switches which can reconfigure in 8~ms.

The remaining tables (\cref{tab:talos-64-cost} - \cref{tab:talos-72-dc-cost}) provide a detailed accounting of the switches used in our various topologies, with switches attributed to the particular topology adaptations or forms of resiliency they are used to provide. Costs are also provided to help show which forms of adaptation are the most expensive. Total costs in these tables reflect switches only; our cost comparisons in \cref{sec:cost-comparison} reflect the total cost including transceivers.

\begin{table}
\centering
\small
\begin{tabular}{|l|c|c|c|c|l|}
\hline
\multirow{2}{*}{Category} & \multicolumn{4}{l|}{Switches per GPU} & \multirow{2}{1.12cm}{Est. Cost per GPU} \\ \cline{2-5}
                   & TP & DP & PP & EP &  \\ \hline
Topo. selection    & \multicolumn{4}{c|}{16 1\by4}  & \$1120         \\ \hline
TP 4 - 8           & 1 2\by2  & 2 2\by2 & & & \$150          \\ \hline
TP 8 - 16  & 0.5 2\by2 & & & & \$25 \\ \hline
PP 4 - 2   & & 2 2\by 2 & & & \$100 \\ \hline
EP 8 - 16          & & & & 2 2\by2 & \$100 \\ \hline\hline
Total & \multicolumn{4}{l|}{16 1\by 4 + 9.5 2\by2} & \$1495 \\ \hline
\end{tabular}
\caption{Switch and cost breakdown for rack-scale clusters without resiliency.}
\label{tab:talos-64-cost}
\end{table}

\begin{table}
\centering
\small
\setlength{\tabcolsep}{4pt}
\begin{tabular}{|l|c|c|c|c|l|}
\hline
\multirow{2}{*}{Category} & \multicolumn{4}{l|}{Switches per GPU} & \multirow{2}{1.12cm}{Est. Cost per GPU} \\ \cline{2-5}
                   & TP & DP & PP & EP &  \\ \hline
Topo. selection    & \multicolumn{4}{c|}{16 1\by4}  & \$1120         \\ \hline
TP resiliency      & \makecell{14.2 1\by2 \\ 1.8 1\by4} & & & & \$437.33\\ \hline
\makecell[l]{PP resiliency \\ (144 GPU only)} & & & \makecell{8 1\by2 \\ 0.9 2\by2} & & \$220.44 \\ \hline
TP 4 $\leftrightarrow$ 8           & 2.7 2\by2 & 4 2\by 2 & & & \$316.67 \\ \hline
TP 8 $\leftrightarrow$ 16          & 0.9 2\by2 &          & & & \$44.44 \\ \hline
PP 4 $\leftrightarrow$ 2           &         & 2 2\by 2 & & & \$100 \\ \hline
EP 8 $\leftrightarrow$ 16          & & & & 2 2\by2 & \$100 \\ \hline\hline
Total (72 GPU)  & \multicolumn{4}{l|}{17.8 1\by 4 + 14.2 1\by2 + 13.6 2\by2} & \$2135.11 \\ \hline
Total (144 GPU) & \multicolumn{4}{l|}{17.8 1\by 4 + 22.2 1\by2 + 14.4 2\by2} & \$2355.55 \\ \hline
\end{tabular}
\caption{Switch count and cost breakdown for rack-scale clusters with resiliency.}
\label{tab:talos-72-cost}
\end{table}

\begin{table}
\centering
\small
\setlength{\tabcolsep}{4pt}
\begin{tabular}{|l|c|c|c|c|l|}
\hline
\multirow{2}{*}{Category} & \multicolumn{4}{l|}{Switches per GPU} & \multirow{2}{1.12cm}{Est. Cost per GPU} \\ \cline{2-5}
                   & TP & DP & PP & EP &  \\ \hline
Topo. Selection    & \multicolumn{4}{c|}{16 1\by4}  & \$1120         \\ \hline
TP 4 $\leftrightarrow$ 8           & 1   2\by2 & 0.5 2\by2 & & & \$75          \\ \hline
TP 8 $\leftrightarrow$ 16          & 0.5 2\by2 & 0.5 2\by2 & & & \$50 \\ \hline
PP 8 $\leftrightarrow$ 4           &           & 0.5 2\by2 & & & \$25 \\ \hline
EP 16 $\leftrightarrow$ 32         & & & & 2 2\by2 & \$100 \\ \hline
EP 32 $\leftrightarrow$ 64         & & & & 2 2\by2 & \$100 \\ \hline
Resiliency      & & \multicolumn{2}{c|}{24 1\by2} & & \$528 \\ \hline
Total & \multicolumn{4}{l|}{16 1\by 4 + 24 1\by2 + 7 2\by2} & \$1998 \\ \hline
\end{tabular}
\caption{Switch count and cost breakdown for multi-rack clusters with rack resiliency only}
\label{tab:talos-64-dc-cost}
\end{table}

\begin{table}
\centering
\small
\setlength{\tabcolsep}{3.5pt}
\begin{tabular}{|l|c|c|c|c|l|}
\hline
\multirow{2}{*}{Category} & \multicolumn{4}{l|}{Switches per GPU} & \multirow{2}{1.12cm}{Est. Cost per GPU} \\ \cline{2-5}
                   & TP & DP & PP & EP &  \\ \hline
Topo. selection    & \multicolumn{4}{c|}{16 1\by4}  & \$1120         \\ \hline
TP Resiliency      & \makecell{14.2 1\by2 \\ 1.8 1\by4} & & & & \$437.32\\ \hline
TP 4 $\leftrightarrow$ 8           & 2.7 2\by2 & 0.5 2\by2 & & & \$158.33 \\ \hline
TP 8 $\leftrightarrow$ 16          & 0.9 2\by2 & 0.5 2\by2 & & & \$69.44 \\ \hline
PP 8 $\leftrightarrow$ 4           &           & 0.5 2\by2 & & & \$25 \\ \hline
EP 16 $\leftrightarrow$ 32          & & & & 2 2\by2 & \$100 \\ \hline
EP 32 $\leftrightarrow$ 64          & & & & 2 2\by2 & \$100 \\ \hline\hline
\makecell{DP+PP Resil. \\ (node only)} & & \multicolumn{2}{c|}{\makecell{24 1\by2 \\ 0.7 2\by2}} & & \$561.33 \\ \hline
\makecell{DP+PP Resil. \\ (node + rack)}      & & \multicolumn{2}{c|}{\makecell{24 1\by4 \\ 1.3 2\by2}} & & \$1713.33 \\ \hline\hline
\makecell[l]{Total (node \\ resilience only)} & \multicolumn{4}{l|}{17.8 1\by4 + 38.2 1\by2 + 9.7 2\by2} & \$2571.42  \\ \hline
\makecell[l]{Total (node + \\ rack resilience)} & \multicolumn{4}{l|}{41.8 1\by4 + 14.2 1\by2 + 10.3 2\by2} & \$3723.42  \\ \hline
\end{tabular}
\caption{Switch count and cost breakdown for datacenter-scale clusters with node-level resiliency}
\label{tab:talos-72-dc-cost}
\end{table}
\begin{table*}[t!]
    \centering
    \begin{tabular}{c|l|lccc}
        \toprule 
        \textbf{Figure} & \textbf{Architecture/Model} & \textbf{Parallelism Configuration} & \textbf{Seq Len} & \textbf{Batch} & \textbf{Num $\mu$B} \\
        \midrule
        \multirow{5}{*}{Fig. 5} & Llama3-8B & TP=4, DP=4, PP=4 & 8196 & 256 & 16 \\
        & Llama3-70B & TP=4, DP=4, PP=4 & 8196 & 256 & 16 \\
        \cmidrule{2-6}
        & Mixtral-7x8B & \small Attn(PP4, TP1, DP16) MoE(PP4, TP1, EP8, DP2) & 8196 & 256 & 16 \\
        & Mixtral-22B & \small Attn(PP4, TP1, DP16) MoE(PP4, TP1, EP8, DP2) & 8196 & 256 & 16 \\
        \cmidrule{2-6}
        & Qwen2-57B-A14B & \small Attn(PP4, TP1, DP16) MoE(PP4, TP1, EP16, DP1) & 16384 & 64 & 16 \\
        \midrule
        \multirow{2}{*}{Fig. 6} & \multirow{2}{*}{Llama-4 Maverick} & \small Dense Layers: TP=8, PP=8, DP=16 & \multirow{2}{*}{4096} & \multirow{2}{*}{1024} & \multirow{2}{*}{16} \\
        & & \small MoE Layers: TP=1, EP=32, DP=4, PP=8 & & & \\
        \bottomrule
    \end{tabular}
        \caption{Model and Parallelism Configurations used in Evaluation Figures 5 and 6.}
    \label{tab:model-configuration}
\end{table*}

\section{Resiliency Analysis}
\label{app:resiliency}

We formalize the resiliency properties of our datacenter-scale deployment, combining both rack-scale and node-scale resiliency.

We first ask the following question: "Given a certain number of failed GPUs in a cluster $p$, what is the chance that the data center scale topology remains non-degraded?". To answer, we consider the failure probability in four domains: a node with 8 GPUs, a node-level resilient rack, a rack-level resilient group of 9 racks, and a whole datacenter. A node fails with at least one faulty GPU, while a rack fails with at least faulty nodes, a rack-level group with at least two faulty racks, and the data center is degraded if at least one rack-level group fails. Simple math shows that the probability to remain operational of a single rack-resilient group with the total of 0.1\% faulty GPUs is 0.017\%.
In a datacenter with 1024 active GPUs (2 resilient groups) the probability to fully recreate a pristine topology is 99.9\%. However, larger deployments with 32,768 active GPUs the chance is 98.9\%. 

At this scale, a \projname topology contains hundreds of thousands of low-radix switches. Therefore, it is important to understand to what degree switch failures are expected. Currently, switch manufacturers sell low-radix switches with quoted minimum lifetimes of at least 10 billion cycles~\cite{glsun-magnetic-switch}, or even claiming no cycle limit for one MEMS-based switch~\cite{sercalo-switch}. For a topology-selection switch cycled 10 times every second, reaching 10 billion cycles would take over 31 years, showing that switches with sufficient lifetimes to be practical are already available.

At the same time, early failures may still be possible. Because these switches have not been deployed at scale, statistics for mean time between failures (MTBF) are not available taking into account the conditions of a realistic datacenter deployment. Instead, we calculate what MTBF would be needed to produce a failure rate of 0.1\%, amortized over all GPUs, over the course of one year. On average, our most resilient topology contains approximately 65 switches per node. This means we can accept one switch out of 65,000 failing over the course of one year, equivalent to a MTBF of 569 million hours.
    
We note that even if a pristine logical topology cannot be perfectly recreated, the remaining topology can continue to be used, often with only minor performance degradation, such as a single missing data-parallel replica or slower data-parallel \allreduce for one pipeline stage.

\section{Training configurations}
\label{app:eval}
\Cref{tab:model-configuration} shows the configuration parameters used to generate our evaluation traces.

\section{MoE  analysis}
\label{sec:appendix:moe}

\Cref{tab:moe-slowdown-analysis} shows the performance of MoE collectives under different configurations to analyze the effect of the expander topology.

\Cref{tab:seqlen} compares the performance of MoE models normalized by the switched network performance, for different sequence lengths.

\begin{table}[t]
    \centering
\begin{tabular}{c|c}
\toprule
Config & Iteration Time [s]\\
\midrule
\projname Recorded MoE &  209.04\\
\projname Uniform MoE  &  205.39\\ 
Fully-connected Uniform MoE  & 171.89\\
\bottomrule
\end{tabular}
    \caption{MoE network performance analysis.}
    \label{tab:moe-slowdown-analysis}
    \vspace{-1em}
\end{table}

\begin{table}[t]
\begin{tabular}{c|ccc}
\toprule
\multicolumn{1}{c|}{\multirow{2}{*}{Model}} & \multicolumn{3}{c}{Sequence Length} \\
\multicolumn{1}{c|}{}                       & 4k         & 8k        & 16k        \\
\midrule
Qwen 57B                                   &   1.43        & 1.34      &   1.29      \\
Mixtral 7x8B                               &  1.06& 1.04  &  1.04  \\
Mixtral 22B                                &   1.05        &  1.04   & 1.04 \\
\bottomrule
\end{tabular}
\caption{Relative training performance across different sequence lengths normalized to the performance of packet switch.}
\label{tab:seqlen}
\end{table}

\end{document}